\documentclass[sigconf,nonacm]{acmart}
\AtBeginDocument{%
  }

\copyrightyear{2026}
\acmYear{2026}
\setcopyright{cc}
\setcctype{by}
\acmConference[UIST '26]{The 39th Annual ACM Symposium on User Interface Software and Technology}{November 02--05, 2026}{Detroit, MI, USA}
\acmBooktitle{The 39th Annual ACM Symposium on User Interface Software and Technology (UIST '26), November 02--05, 2026, Detroit, MI, USA}
\acmDOI{10.1145/3830398.3830523}
\acmISBN{979-8-4007-2856-3/2026/11}

\newcommand{\redacted}[1]{\textbf{REDACTED}}

\usepackage{booktabs}

\usepackage{multirow}
\usepackage{float}
\usepackage{graphicx}
\usepackage{subcaption}
\usepackage{makecell}
\usepackage{xcolor}      

\newenvironment{promptbox}
  {\par\footnotesize\ttfamily\raggedright\setlength{\parindent}{0pt}\frenchspacing}
  {\par}
\newenvironment{UIST}
  {\begingroup\color{black}}  
  {\endgroup}              

\newcommand{\revise}[1]{\textcolor{black}{#1}}

\newcommand{\italquote}[1]{\begin{quote}``\textit{#1}''\end{quote}}

\hypersetup{colorlinks=true, allcolors=blue}
\begin{document}

\title{Generative Experiences for Digital Mental Health Interventions: Evidence from a Randomized Study}


\author{Ananya Bhattacharjee}
\affiliation{%
  \institution{Stanford University}
  \city{Stanford}
  \state{California}
  \country{USA}
}
\email{ananyabh@stanford.edu}

\author{Michael Liut}
\affiliation{%
  \institution{University of Toronto Mississauga}
  \city{Mississauga}
  \state{Ontario}
  \country{Canada}
}
\email{michael.liut@utoronto.ca}

\author{Matthew Jörke}
\affiliation{%
  \institution{Stanford University}
  \city{Stanford}
  \state{California}
  \country{USA}
}
\email{joerke@stanford.edu}

\author{Diyi Yang}
\authornote{Equal contributions}
\affiliation{%
  \institution{Stanford University}
  \city{Stanford}
  \state{California}
  \country{USA}
}
\email{diyiy@stanford.edu}

\author{Emma Brunskill}
    \authornotemark[1]
\affiliation{%
  \institution{Stanford University}
  \city{Stanford}
  \state{California}
  \country{USA}
}
\email{ebrun@cs.stanford.edu}



\begin{abstract}

Digital mental health (DMH) tools have extensively explored personalization of interventions to users’ needs and contexts. However, this personalization often targets what support is provided, not how it is experienced. Even well-matched content can fail when the interaction format misaligns with how someone can engage. We introduce generative experience as \revise{an approach to DMH support}, where the intervention experience is composed at runtime. We instantiate this in GUIDE, a system that generates personalized intervention content and multimodal interaction structure through rubric-guided generation of modular components. In a preregistered study with N=237 participants, GUIDE significantly reduced stress  (p=.02) and improved user experience (p=.04) compared to an LLM-based cognitive restructuring control. GUIDE also supported diverse forms of reflection and action through varied interaction flows, while revealing tensions around personalization across the interaction sequence. This work lays the foundation for interventions that dynamically shape how support is experienced and enacted in digital settings.


\end{abstract}


\begin{teaserfigure}
\centering
  \includegraphics[width=\textwidth]{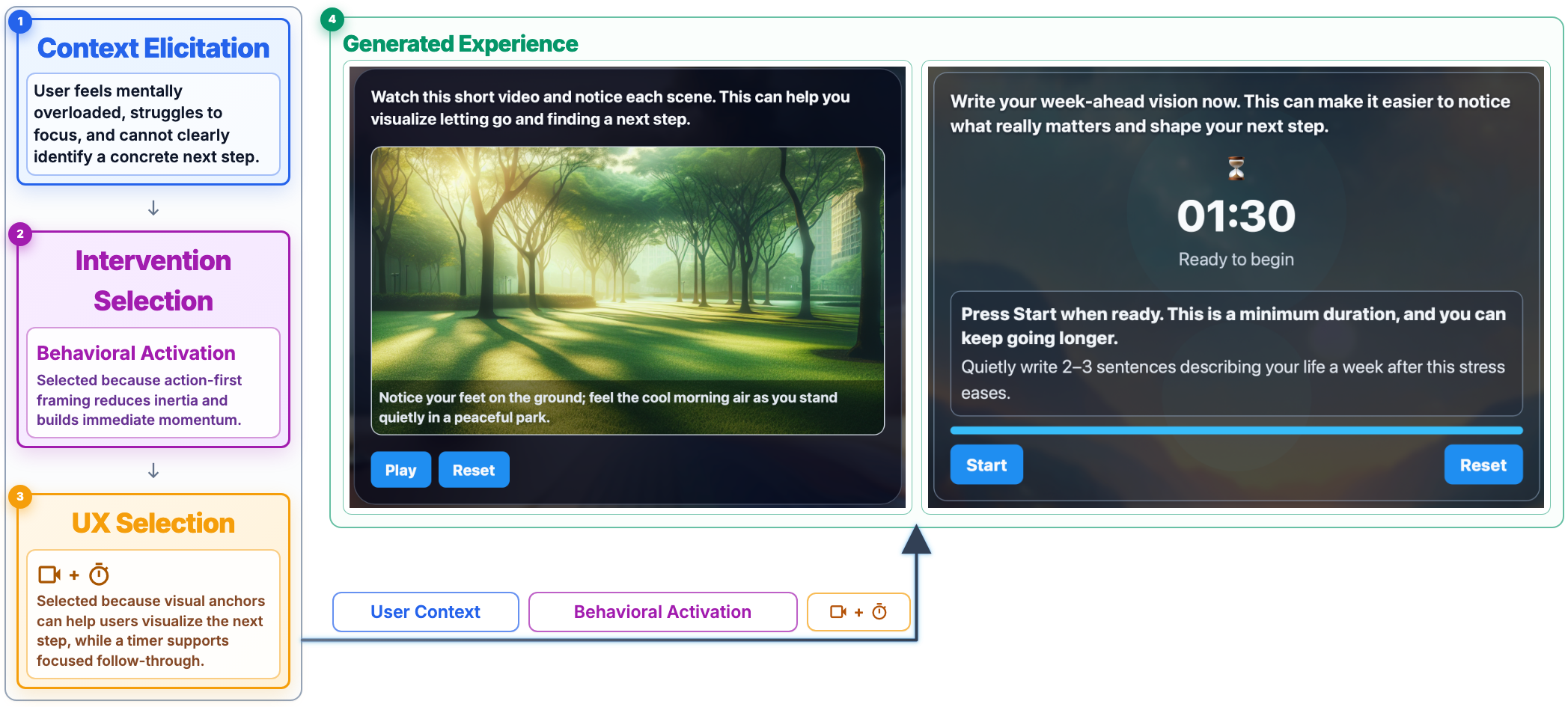}
  \caption{GUIDE generates intervention experiences at runtime by composing interaction structures from user context and selected interventions. The system elicits context, selects an intervention, and constructs a multimodal experience using elements such as visual guidance and timed activities. The interfaces shown are illustrative snippets of generated UX.
  }
  \label{fig:teaser}
\end{teaserfigure}


\maketitle


\section{Introduction}


Psychological interventions are often structured as activities such as reflecting on a situation \cite{bhattacharjee2024exploring, bono2013building}, examining thoughts \cite{sharma2023cognitive, beck1979cognitive}, or taking small actions \cite{meyerhoff2024small}. Consider a simple activity where a person revisits a stressful situation from a third-person perspective. This could be carried out by writing a short reflection, following guided instructions, or listening to a brief audio message. It could unfold as a quick one-step activity or a longer sequence. Even when the underlying intervention is the same, these differences shape the \textit{experience} of the activity and influence whether it supports the intended mental health outcomes \cite{liu2023moving, slovak2024hci}.

In practice, clinicians and other support providers actively adapt how an intervention is delivered based on contextual factors such as a person’s mood, energy, or surroundings \cite{stiles1998responsiveness, chorpita2005modularity, bhattacharjee2023investigating, bhattacharjee2025investigating, kornfield2020energy}. When someone feels overwhelmed, an activity may be simplified into a few guided steps or a short audio-based experience. In other situations, it may expand into a longer, more reflective exercise. These adjustments shape how the activity unfolds and whether it can be carried out as intended.

Digital mental health (DMH) tools, such as web-based programs \cite{sharma2023cognitive}, AI chatbots for stress support \cite{meyerhoff2024small}, and mood tracking apps \cite{schueller2021understanding}, aim to provide this type of support at scale. Prior work has made substantial progress in adapting intervention content to users’ needs, preferences, and contexts \cite{bhattacharjee2023investigating, bhattacharjee2025perfectly, sharma2023cognitive, liu2024compeer}. However, prior work has largely neglected adapting the experiential dimension of DMH interventions to the user \cite{liu2023moving, slovak2024hci}. Most systems rely on predefined UI/UX templates or fixed interaction workflows, where the structure of the interface is determined at design time and content adapts only within the constraints of that interaction. As a result, both the interaction format and the range of intervention activities are constrained, limiting support to what can be expressed within a fixed interface.

This can create mismatches between a user’s situation and how support is carried out. Even when the selected intervention is appropriate, a fixed interaction format may not support the reflective or behavioral process that intervention requires. For example, a system designed for cognitive restructuring typically supports only restructuring activities \cite{sharma2024facilitating}, with limited ability to shift to other forms of support such as guided reflection, breathing exercises, or multimodal activities.

\begin{UIST}
    
In this work, we refer to generative experience as the ability of a system to dynamically construct a personalized intervention experience: both the support a user receives and the interaction through which that support is enacted. This framing treats the experience itself, and not only the intervention content, as an object of generation. The same underlying intervention can be realized through different experiences, and which experience is constructed shapes how the user engages with it. For example, revisiting a stressful situation from a third-person perspective could be delivered as a short audio-guided activity or a longer structured writing task. Personalization is driven by two inputs: 1) the user's elicited context, including the nature of the stressor, surrounding circumstances, and perceived controllability, and 2) the demands of the selected intervention. The experience is composed to fit both, spanning modality, structure, and sequence.
\end{UIST}

In doing so, we shift the problem from selecting only the right intervention to both selecting and generating the interaction through which that intervention takes form. Psychological interventions require users to carry out processes such as reflection, cognitive reframing, or action planning, and the interface shapes how those processes unfold by structuring what users do, in what sequence, and with what support \cite{bhattacharjee2024understanding, slovak2024hci}. We posit that when the experience is better aligned with both the demands of the intervention and the user's context, users may be more likely to carry out the intended psychological process, increasing the likelihood of proximal benefits such as reflection or reappraisal \cite{nahum2016just, klasnja2015microrandomized}, which can in turn contribute to improved mental health outcomes.

We instantiate this \revise{approach} in \textbf{GUIDE} (\textbf{G}enerative \textbf{U}I for \textbf{I}nter-active \textbf{D}igital \textbf{E}xperiences), a system for generating personalized intervention experiences at runtime. 
\revise{Following \citet{slovak2024hci}, we position GUIDE as an intervention system, an end-to-end ensemble of components (intervention selection, rubric-guided evaluation, modular UX composition) targeting a psychological outcome.}
GUIDE treats both the intervention and the user interface through which it is enacted as objects of generation. Building on prior work in adaptive interface generation \cite{tyler1986adaptive, gajos2004supple, chen2025generative} and generative systems that synthesize interaction flows from goals and constraints \cite{vaithilingam2024dynavis, li2025sketch2code, si2025design2code}, GUIDE formulates DMH support as a structured composition problem over intervention strategies and interaction forms. The system begins by eliciting user context through a short guided interaction, which can incorporate multimodal input such as text or voice. It then generates multiple candidate interventions, represented as structured activity sequences, and selects among them using rubric-guided evaluation. Conditioned on the selected intervention and user context, GUIDE generates multiple candidate interaction realizations by composing parameterized modules from a bounded design space, including elements such as input fields, audio guidance, timed sequences, and visual components, and again selects among them using rubric-guided evaluation. 

This two-stage generative design means the same intervention can be enacted through different interaction structures, and different intervention strategies can be paired with different interaction forms — producing a combinatorial design space instead of a fixed interface. By combining modular multimodal composition with rubric-guided candidate selection at both stages, GUIDE moves beyond single-pass generation to dynamically assemble intervention experiences tailored to each user context.

We evaluated our approach through a preregistered between-participant study of a single-session stress management intervention with 237 participants. 
\revise{Our evaluation asks whether the integrated system improves outcomes relative to a strong alternative system: a well-validated LLM-based cognitive restructuring approach \cite{sharma2023cognitive, sharma2024facilitating}}.
Across this evaluation, GUIDE produced greater immediate stress reduction ($p = .02$) and better overall user experience ($p = .04$) than the control. GUIDE also improved related perceptions, such as enjoyment and willingness to use similar activities again. Further analyses showed that GUIDE generated a diverse range of support techniques and interaction experiences, which participants described as helping them shift perspectives on stressful situations and make small progress. At the same time, the findings revealed important tensions in how personalization was experienced across the interaction. 

Our contributions include:
\begin{itemize}    
\item \revise{Conceptualizing generative experience as an approach to designing DMH intervention systems},
    \item Designing GUIDE, a system for generating personalized intervention experiences at runtime, and
    \item Empirically evaluating this \revise{approach} in a single-session stress management study with $N=237$ participants against an LLM-based cognitive restructuring control.
\end{itemize}

\section{Related Work}

We review two strands of prior work: personalization and interaction in DMH systems, and adaptive user experience generation.

\subsection{Personalization and Interaction in DMH Tools}

Earlier DMH systems have personalized intervention content through approaches such as scripted responses \cite{bhattacharjee2022kind}, decision trees \cite{abd2019overview}, crowdsourced responses \cite{morris2014crowd, morris2015efficacy, smith2021effective}, and NLP-based conversational agents \cite{fitzpatrick2017delivering, meyerhoff2024small}. Recent advances in generative AI have significantly expanded this capability by enabling systems to process open-ended input and generate responses that adapt to the user’s context~\cite{sharma2023cognitive, sharma2024facilitating, jo2023understanding, jo2024understanding, bhattacharjee2025perfectly, liu2024compeer, lo2025d, fang2025social}.
For example, LLM-based tools have been developed to generate cognitive reframing suggestions tailored to the specific difficulties described by users \cite{sharma2023cognitive, sharma2024facilitating}.  Generative models have also been used to create personalized narratives that guide users through reflective scenarios \cite{bhattacharjee2025perfectly} or facilitate peer support-style conversations around personal challenges \cite{liu2024compeer}.

While dynamic generation of DMH content has advanced considerably, far less attention has been given to dynamically generating the experience through which interventions are supported. 
Some recent work has introduced limited variance in interaction experience \cite{bhattacharjee2024understanding, song2025exploreself, guo2025exploring}. For instance, ExploreSelf supports reflective writing through dynamically generated themes and summaries \cite{song2025exploreself}, and \citet{bhattacharjee2024understanding} has generated structured strategies and prompts within task-oriented interfaces. However, in these systems the overall modality and activity structure remain predetermined, with generative models primarily used to populate or guide elements within a fixed interface. 

Related work has also explored multimodal interaction in DMH support \cite{wang2025and, lim2024exploring, bao2025milo, balban2023brief, silverstone2016complex, vowels2025evaluating}.  Voice-based interaction has enabled engagement when typing is difficult and supported guided interaction through spoken instructions or conversational input \cite{vowels2025evaluating, berube2021voice}. Visual elements such as images and videos have made activities easier to follow and more engaging by illustrating concepts and scenarios \cite{harshbarger2021challenges, galmarini2024effectiveness}. Timer-based elements have been used to pace activities such as breathing or relaxation, helping users regulate engagement and sustain attention \cite{wennberg2018effectiveness, balban2023brief}. Other systems have incorporated modalities such as touch, avatars, and expressive cues to enhance engagement and social expressiveness \cite{bao2025milo, jin2025don, silverstone2016complex, jorke2025bloom}. Yet these modalities are typically introduced within predefined interaction structures, extending fixed activities rather than supporting dynamic generation of the intervention experience.

Altogether, prior work has made substantial progress in personalizing intervention content based on user context, but the interaction experience is still typically fixed in advance. In this work, we build on these advances by enabling generative experience, where the interaction experience through which an intervention is enacted is constructed at runtime.





\subsection{Adaptive User Interface Generation}

HCI research has long recognized the importance of adaptive user interfaces, with early approaches relying on predefined rules and standards to adjust elements such as layout, widgets, and interaction based on user, device, and task characteristics \cite{tomlinson2007dreaming, maybury1998intelligent, tyler1986adaptive, horvitz1999principles, gajos2004supple, ponnekanti2001icrafter}. Systems such as SUPPLE \cite{gajos2004supple, gajos2007automatically} and related work on activity-oriented interfaces demonstrate these capabilities \cite{smith2003groupbar, houben2013activity, park2024coexplorer}, but remain limited in scalability as adaptation is largely rule-based and tied to predefined mappings or variants.

Model-Based User Interface (MBUI) development aims to manage the complexity of building adaptive interfaces by separating application logic from interface design through structured models \cite{cao2025generative, myers1995user, szekely1992facilitating}. These models represent tasks and domain data and map abstract interaction descriptions to concrete interface elements, allowing developers to specify interaction requirements at a higher level of abstraction \cite{klemmer2004tangible}. Specification-based UI generation extends this approach by defining interaction requirements using structured primitives and translating them into interface elements through predefined mappings \cite{puerta1998towards, nichols2002generating, nichols2004improving, vaithilingam2019bespoke, vaithilingam2024dynavis}. For example, instead of specifying a fixed layout, a developer may declare that multiple related inputs are needed, and the system determines whether to present them as a form, multi-step flow, or layout adapted to device constraints. While these approaches support greater flexibility, they still depend on predefined mappings that limit how interaction structures can vary at runtime.

Recent advances in generative AI have renewed interest in adaptive UI and UX generation by making it possible to translate natural language input into functional code, interface structures, and interactive artifacts. Many widely used systems such as GPT and Claude can already produce simple interfaces from model-generated code, while recent research has extended these capabilities by generating interfaces from natural language descriptions \cite{laurenccon2024unlocking}, screenshots \cite{si2025design2code}, and sketches \cite{li2025sketch2code}. As these systems move beyond one-shot generation, they increasingly rely on modular pipelines in which different components handle distinct stages such as interpreting user intent, representing interface structure, and rendering the final interface \cite{wang2024promptcharm, petridis2024constitutionmaker}. This modular organization is especially visible in recent work that builds on ideas from MBUI and specification-based UI generation. These approaches introduce intermediate representations of interface structure that specify what controls, visual elements, and interaction flows should be instantiated for a given task \cite{chen2025generative, luera2026survey}. Some systems further improve generation quality by producing multiple candidate interfaces and refining them through rubric-based evaluation and iterative feedback \cite{chen2025generative}.

In short, prior work on adaptive user interfaces has primarily focused on modifying predefined interface structures. Recent generative AI systems expand this by enabling more flexible integration of UX components and interaction flows. This flexibility may be particularly important in DMH settings, where how support is experienced could shape how users work through an activity \cite{slovak2024hci}. We extend this direction by generating the interaction experience itself, first selecting an appropriate intervention, and then constructing how it is structured, sequenced, and enacted at runtime based on user context in a DMH setting. By adapting how an activity unfolds, generative UI may better align with the demands of the intervention and the user’s situation.


\section{Design}


GUIDE  was iteratively developed through multiple rounds of internal design and expert feedback. Members of the research team have prior experience designing DMH interventions, both with and without AI, and have published in leading HCI, psychology, and AI venues. We also conducted semi-structured consultation sessions with six experts, which informed iterative refinements. Additional details about expert backgrounds and consultation sessions are provided in Appendix \ref{app:expert}.

\subsection{Goal}
GUIDE is designed to generate personalized intervention experiences tailored to a user’s situation. The system aims to identify an appropriate intervention for the user’s context and construct an interaction experience that presents the intervention. It also aims to ensure that both the intervention and the interaction experience maintain quality and alignment with established psychology and UX principles.

GUIDE was designed to deliver a single-session intervention (SSI) \cite{schleider2017little, schleider2020acceptability}, a structured, standalone activity intended to provide focused support and produce immediate changes in targeted outcomes. 
SSIs are designed to offer brief support within a short time window, typically 10–20 minutes, and have been shown to promote reflection and cognitive reappraisal while remaining accessible to a wide range of users. 
Prior work has shown that such interventions can help individuals manage negative thoughts, reduce stress, and improve anxiety and depression symptoms across both clinical and non-clinical populations \cite{sharma2023cognitive, schleider2017little, schleider2020acceptability, bhattacharjee2024exploring, bhattacharjee2025perfectly}.

\subsection{System Implementation}
\begin{figure*}
    \centering
    \includegraphics[width=0.9\linewidth]{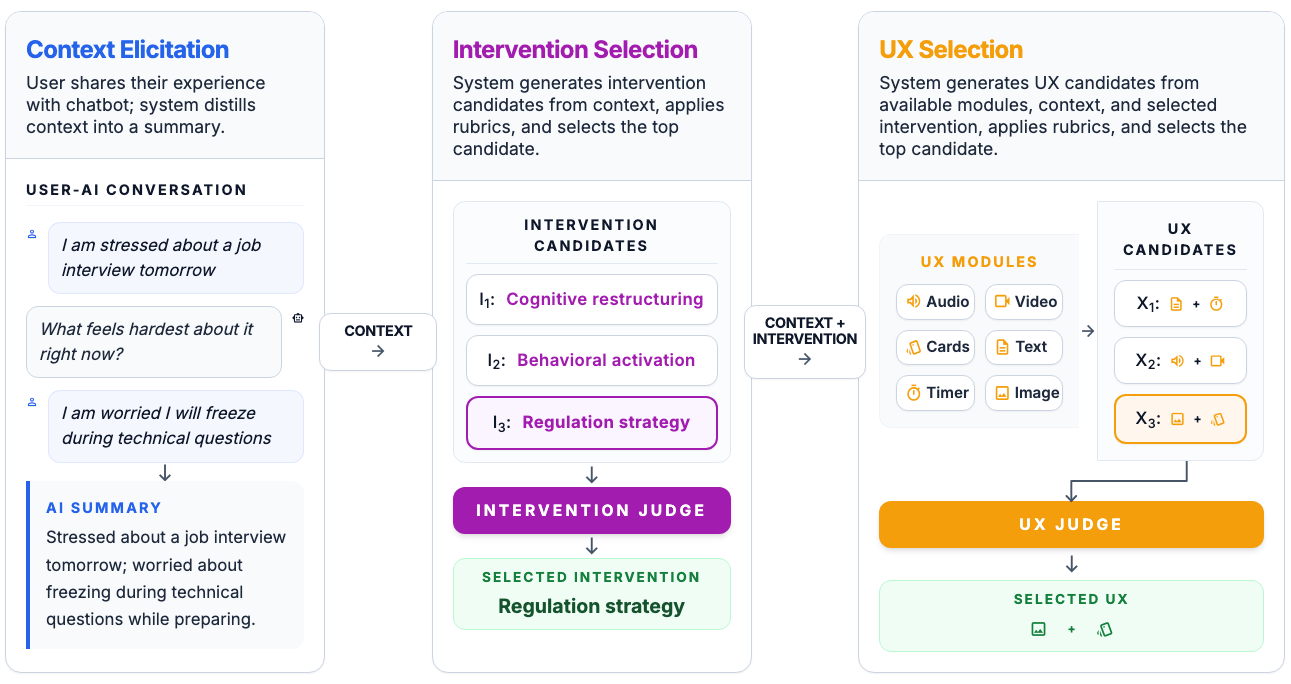}
    \caption{System overview of GUIDE. The system first elicits user context through a structured conversation and produces a concise summary. It then generates multiple candidate interventions from this context and selects one using rubric-guided evaluation. Given user context and the selected intervention, the system composes candidate UX realizations from a set of interaction modules (e.g., audio, text, timer) and selects the final experience through a second rubric-based selection process.}
    \label{fig:system_diagram}
\end{figure*}

GUIDE has three major components: (1) context elicitation, (2) intervention selection, and (3) UX selection. Figure \ref{fig:system_diagram} provides a high-level overview, and we describe each component below. \revise{The project website, which includes code and additional details, is available at \textcolor{blue}{\url{https://ananya-bhattacharjee.github.io/guide/}}.}

\noindentparagraph{1. Context Elicitation:} To generate contextually appropriate support, the system began with a brief guided elicitation phase with the user to capture the user’s stress context. The elicitation used a small set of structured prompts grounded in prior work on stress and DMH support to capture key psychological and situational dimensions relevant to stress experiences \cite{bhattacharjee2024exploring, skeggs2025micro, bhattacharjee2026user, o2018suddenly}. Users could respond either by typing or by using voice input, and prompts could also be read aloud through AI generated voice playback so that both text and voice interaction were supported. The elicitation included five guided questions covering situation $\rightarrow$ difficulty $\rightarrow$ impact $\rightarrow$ sense of control $\rightarrow$ current context.



Each question was phrased to remain generalizable across short term, long term, and intermittent stress experiences.  The interaction was implemented as a state-based flow over five questions. 
After each response, an LLM checked whether the information sufficiently captured the intended dimension and issued a short clarification prompt if needed. To avoid overwhelming users, at most one clarification follow-up was asked per question before proceeding. AI responses also included brief acknowledgments to maintain a conversational tone while guiding the interaction forward.

After the elicitation phase, the system generated a short two-paragraph summary of the user's situation from the conversation. Prior work suggests that summaries can help make complex personal situations easier to interpret, support reflection, and provide a shared representation that users can inspect and revise~\cite{baumer2015reflective, kim2024mindfuldiary}. 
They could either edit the text manually or request further revisions. 
We denote the final context used for downstream intervention and UX generation by $C$, which includes both the information gathered through the five guided questions and the summary of the situation.




\noindentparagraph{2. Intervention Selection:} 
GUIDE was instructed to generate brief, structured intervention activities informed by principles from Cognitive Behavioral Therapy (CBT) \cite{beck2020cognitive}. CBT covers a broad range of support behaviors (e.g., cognitive restructuring, behavioral activation, problem solving \cite{beck2020cognitive}) and is one of the most widely used families of psychological interventions \cite{salkovskis2023effective}. One aspect of generativeness arises from this breadth, as GUIDE was instructed to draw from any CBT-informed web-based activities rather than being limited to a pre-defined set. 


During initial testing we observed that when LLMs were asked to propose contextual activities, 
they tended to focus heavily on reflective emotional processing, despite CBT encompassing a broader range of possible support behaviors. Based on consultations with experts, we encouraged diversity in generated activities by using few-shot prompting and instructing the model to produce a set of candidate interventions $\mathcal{I}(C) = \{I_1, I_2, \dots, I_n\}$, where each $I_k$ represents a short structured activity implementing an intervention strategy. \revise{We set $n = 3$ in our implementation. We settled on this value during iterative development to balance candidate diversity against the generation and rubric-evaluation latency each additional candidate adds in a live interactive setting.} 


\begin{UIST}    
The generation prompt operationalizes CBT grounding along two complementary strategies: thought-focused and action-focused interventions \cite{beck2020cognitive}. 
A thought-focused intervention targets a change in understanding through one of seven suggested techniques (reappraisal, attribution rebalancing, distancing, evidence testing, cost-benefit analysis, values clarification, or hypothesis generation) and must conclude with a visible output that records the resulting shift, for instance a reframed interpretation set beside the original thought, or a short list separating evidence for and against a worry. An action-focused intervention centers on a behavioral technique drawn from six options (graded exposure, micro action planning, implementation intentions, problem decomposition, assumption testing, or communication rehearsal) and requires the user to enact the core unit of the skill, for instance rehearsing aloud a difficult conversation, or committing to one specific next step with a concrete cue. The prompt asks each generated intervention to integrate thought-focused and action-focused support into one activity. It requires the selected techniques to be named, central to the activity, practiced during the interaction, and grounded in at least three details from the user’s own account, so that the intervention remains tied to the user’s situation instead of becoming a generic exercise. These technique options are illustrative and not exhaustive, leaving room for the model to compose other CBT-informed interventions.
\end{UIST}


\begin{UIST}
    
To evaluate the generated interventions, we developed a set of eight \textit{intervention judge rubrics} in close consultation with experts. These rubrics were refined through multiple iterations to capture both psychological principles and practical considerations relevant to brief SSIs \cite{persson2025design, paredes2014poptherapy, gottfredson2015standards, mohr2014behavioral, yardley2015person}. 
Specifically, the rubrics are (1) Narrative Flow, (2) Small Progress, (3) Safe Sequencing, (4) Explicit Alignment with Psychological Principles, (5) Specificity, (6) Non-retrievability, (7) Everyday Feasibility, and (8) Understandability. Each rubric was specified through a short definition, a contrasting pair of high- and low-quality examples, a set of concrete checks, and anchored scale points to calibrate scoring. Each rubric $r \in \mathcal{R}_I$ was scored on a 1--5 scale by an LLM judge given the candidate intervention and user context $C$, with higher scores indicating better alignment; all criteria were weighted equally. \revise{To support discrimination among closely matched candidates and reduce order bias, the judge compared all candidates side by side under each criterion before scoring (a listwise evaluation~\cite{li2024llms}) and grounded every rating in concrete details drawn from the plan text}. The candidates were evaluated using these rubrics through an LLM-as-a-judge approach \cite{li2024llms, chandra2025lived}. Additional details about these rubrics are provided in Appendix~\ref{app:int-judge-rubric}, with full prompts in Appendix~\ref{app:rubric-intervention}.

\end{UIST}

The overall score for an intervention was computed as
\begin{equation}
S_I(I_k, C) = \sum_{r \in \mathcal{R}_I} R_{I,r}(I_k, C).
\end{equation}
The system then selected the highest scoring intervention
\begin{equation}
I^* = \arg\max_{I_k \in \mathcal{I}(C)} S_I(I_k, C).
\end{equation}

We illustrate a hypothetical example of rubric-based intervention selection in Figure~\ref{fig:judge} in Appendix~\ref{app:int-judge-rubric}. 




\noindentparagraph{3. UX Selection:} Our UX generation process drew on literature on specification-based interface generation~\cite{puerta1998towards, vaithilingam2024dynavis}, where interfaces are first described through structured components (i.e., selected intervention $I^*$) and then instantiated into concrete interaction flows. Generativeness in the user experience comes from constructing the interaction at runtime, where primitives can be combined and repeated in many possible orders, resulting in a combinatorial design space rather than a fixed interface. Similar to intervention generation, the system was instructed to generate a set of candidate interaction experiences $\mathcal{X}(I^*, C) = \{X_1, X_2, \dots, X_n\}$. We again set $n = 3$. 

We defined a set of modular interaction components, which were derived from a review of implementations of common web-based SSIs in which users complete a short activity aimed at a mental health outcome \cite{bhattacharjee2025perfectly, sharma2024facilitating, sharma2023cognitive, schleider2017little, schleider2020acceptability, kaveladze2026crowdsourced}. The set consisted of primitives ($\tau$) such as text input, audio message, or timer (see Table~\ref{tab:ux_palette_short} for examples). They belonged to a subset of the four broad interaction types -- text, audio, visual, and temporal (see Figure~\ref{fig:ux_palette}).

Each primitive $\tau$ was instantiated through configurable parameters $\theta$ that determine how it appeared and behaved in context. For example, audio messages included parameters such as script and tone, and timers included duration and associated prompts. Table \ref{tab:ux_palette} illustrates these primitives with configurable parameters and interaction types. Each candidate $X_j$ was represented as a sequence of parameterized interaction elements, $X_j = ((\tau_1, \theta_1), (\tau_2, \theta_2), \dots, (\tau_T, \theta_T))$, where $\tau_t$ denotes the interaction primitive and $\theta_t$ specifies its parameters.

\begin{table}[t]
\centering
\small
\caption{Representative interaction primitives used to construct generated intervention experiences. The full set of primitives is shown in Table~\ref{tab:ux_palette}.}
\label{tab:ux_palette_short}
\setlength{\tabcolsep}{4pt}
\begin{tabular}{@{} p{2cm} p{3.7cm} p{2.2cm} @{}}

\toprule
\textbf{Primitive ($\tau$)} & \textbf{Parameters ($\theta$)} & \textbf{Interaction Type} \\
\midrule

Text Input &
prompt question; response hint; intervention purpose &
Text \\

Choice Input &
prompt question; response options; multiple selection; intervention purpose &
Text \\

Audio Message &
audio script; delivery tone; speaking rate; intervention purpose &
Text/Audio \\

Guided Sequence &
timed cue steps; audio cue script; intervention purpose &
Text/Audio/Temporal \\

Image Display &
image description prompt; intervention purpose &
Text/Visual \\

Timer &
duration; timer text; reflection prompt; intervention purpose &
Text/Temporal \\
\bottomrule

\end{tabular}
\end{table}


To evaluate candidate interaction experiences, we developed a set of seven \textit{UX judge rubrics}. These rubrics capture key aspects of interaction quality, including (1) Intervention-Interface Alignment, (2) Task Efficiency, (3) Usability, (4) Information Clarity, (5) Interaction Satisfaction, (6) Specificity, and (7) Understandability \cite{nielsen1994usability, hartmann2008towards, yardley2015person, brooke1996sus}. \revise{Each criterion is defined with high- and low-quality examples, a checklist, and anchored scale points, and is scored on a 1--5 scale by an LLM judge}.   Candidate experiences were evaluated using these rubrics $\mathcal{R}_X$. \revise{Additional details about these rubrics are provided in Appendix~\ref{app:int-judge-rubric}, with detailed prompts in Appendix~\ref{app:rubric-ux}}. Let $R_{X,r}(X_j, I^*, C)$ denote the score assigned to UX candidate $X_j$ under rubric $r \in \mathcal{R}_X$. The overall score was computed as
\begin{equation} 
S_X(X_j, I^*, C) = \sum_{r \in \mathcal{R}_X} R_{X,r}(X_j, I^*, C).
\end{equation}
The system then selected the highest scoring candidate
\begin{equation} 
X^* = \arg\max_{X_j \in \mathcal{X}(I^*, C)} S_X(X_j, I^*, C).
\end{equation}



\revise{GUIDE used GPT-4.1 with default parameters (temperature = 1) to generate and evaluate both interventions and their corresponding interaction experiences. Multimodal elements were rendered using specialized APIs (DALL·E 3 for images and GPT-4o-mini-based models for speech input and output). The system also included a fallback intervention in case of generation failure, though it was never triggered during the study. Across the pipeline, the stages, interaction primitives, rubrics, and number of candidates (n = 3) were fixed at design time, while the intervention strategy, content, modalities, and interaction sequences were generated at runtime. In our user study (Section~\ref{sec:study}), GUIDE sessions lasted 18.6 minutes on average, including a mean end-to-end generation latency of 52 seconds.}

\begin{UIST}
\subsection{Validating the Generation Pipeline}
\label{sec:pipeline-validation}

As a configuration check, we conducted an ablation with simulated users examining the role of rubric-guided generation in intervention and UX composition (see Appendix~\ref{app:ablation}). The full pipeline with rubric guidance at both stages was most frequently selected as best across simulated contexts, informing our decision to retain rubric guidance in the deployed system.

We also ran a separate evaluation to verify that its selections
aligned with human judgment. We sampled 40 sessions and, for each, recorded the three candidates the system generated at each stage (three interventions produced from the user context, and three UX realizations produced from the context and the selected intervention). Two experts (E5 and E6) then independently selected the best of the three candidates at each stage, viewing the same information the judge received: for interventions, the user context; for UX, the context and the selected intervention. Raters were blind to which candidate the judge had selected. Each expert judged 30 of the 40 sessions per stage, with 20 sessions overlapping.

We measured agreement with Cohen's $\kappa$, treating each selection as a choice among three candidates. The judge showed substantial agreement with the expert raters for both interventions ($\kappa = 0.67$) and UX ($\kappa = 0.69$), approaching the agreement between the raters themselves ($\kappa = 0.75$ and $\kappa = 0.77$, respectively). On sessions where the two raters selected the same best candidate, the judge matched their
consensus in 82.4\% of cases at each stage. Together, these analyses indicate that rubric-guided selection is both beneficial to output quality and aligned
with expert judgment, though agreement is not perfect and the judge validation was limited to two raters over a sampled set of sessions.

\end{UIST}

\section{User Study}
\label{sec:study}

We conducted a pre-registered, between-participant experiment (N=237) in which participants were randomly assigned to either our system or a control condition (Section~\ref{subsub:control}) within a single session. The study was approved by the Institutional Review Boards (IRB) at Stanford University and the University of Toronto. We describe the study methods below.

\subsection{Participants}

Participants were recruited from an undergraduate computer science course at a major North American university through a course announcement. \revise{University students are a relevant population for stress and mental health research \cite{madrid2025digital}, since computing students in particular report high rates of stress, anxiety, and depression \cite{danowitz2018characterizing, chandrasekaran2025spent}.}
Participation from the students was voluntary, and those who completed the study and correctly submitted their identifying information received a 2\% course bonus. Participants were required to be at least 18 years old. Attention check questions were included before and after the intervention, and responses failing these checks were excluded from analysis.

In total, 250 participants completed the activity. Thirteen participants were excluded due to failed attention checks, resulting in a final dataset of 237 participants (Control: $n=115$, CP1–CP115; GUIDE: $n=122$, GP1–GP122). The mean age was 20.78$\pm$1.3 years. Participants identified with multiple genders (178 men, 47 women, 3 non-binary, and 9 undisclosed) and several racial groups (177 Asian, 27 White, 3 African American, 9 mixed race, and 21 undisclosed). \revise{Based on standard score ranges in PSS-10 scale, 29 participants fell in the low stress range (0-13), 163 in the moderate stress range (14-26), and 45 in the high stress range (27-40).}

\begin{table}[t]
\centering
\small
\setlength{\tabcolsep}{4pt}
\renewcommand{\arraystretch}{0.95}
\caption{Outcome measures used in the study.}
\label{tab:measures}
\begin{tabular}{@{} >{\raggedright\arraybackslash}p{2.9cm} p{5.2cm} @{}}

\toprule
\textbf{Outcome} & \textbf{Measure and Scoring} \\
\midrule

\multicolumn{2}{c}{\textbf{Primary outcomes}} \\
\midrule

Stress reduction &
“How stressful is the situation you are thinking about?” measured before and after the activity \cite{bhattacharjee2024exploring}. Rated 1--5; $\textrm{Stress}_{\textrm{pre}} - \textrm{Stress}_{\textrm{post}}$. \\

User experience &
UEQ-8 user experience scale measured after the activity \cite{hinderks2017design}. Mean of 8 items, rescaled to $[-2,2]$. \\

\midrule
\multicolumn{2}{c}{\textbf{Exploratory outcomes}} \\
\midrule

Stress mindset improvement &
8-item stress mindset scale measured before and after the activity \cite{crum2013rethinking}. Rated 0--4; $\textrm{Mindset}_{\textrm{post}} - \textrm{Mindset}_{\textrm{pre}}$. \\

Perceived personalization &
“The suggested activity felt personalized to my specific situation.” (post, 1--5 agreement) \\

Perceived system understanding &
“The system understood my situation and concerns when suggesting the activity.” (post, 1--5 agreement) \\

Perceived reflection of user input &
“The activity reflected information I shared in a way that felt relevant.” (post, 1--5 agreement) \\

Intent to reuse activity &
“I would like to use a similar activity again in the future.” (post, 1--5 agreement) \\

Intent to recommend activity &
“I would recommend this activity to others experiencing stress.” (post, 1--5 agreement) \\

Activity length appropriateness &
“The length of the activity felt appropriate.” (post, 1--5 agreement) \\

Activity enjoyment &
“I enjoyed taking part in this activity.” (post, 1--5 agreement) \\

\bottomrule
\end{tabular}
\end{table}

\subsection{Study Procedure}

The activity was delivered through a web-based interactive system. Upon accessing the system, participants reviewed and provided informed consent and completed pre-activity questions, which indicated that the activity would use AI and technology to provide stress support. Participants were then randomly assigned to a condition, completed the intervention within the system, and answered post-intervention questions. Participation was asynchronous, allowing participants to complete the study at their convenience within a predefined time window. 

Participants were  provided with emergency resources such as crisis text lines and suicide helplines. We did not solicit suicide-related information, although open-ended responses allowed for the unlikely possibility of distress disclosure. Both conditions used OpenAI’s moderation tools along with daily manual review to monitor risk. We had a prior, IRB-approved protocol for responding to participants flagged as at risk of significant mental distress; however, no participants were flagged during the study.



\subsubsection{Design of Control Condition}
\label{subsub:control}

We implemented a control condition adapted from a prior cognitive restructuring system \cite{sharma2023cognitive, sharma2024facilitating}, representing a strong, well-established LLM-based approach. This control was chosen because it reflects a widely used and empirically validated form of CBT support. \revise{It has been deployed at scale through a national mental health nonprofit (Mental Health America) with positive mental health outcomes \cite{sharma2023cognitive, sharma2024facilitating}.} It follows a three-stage workflow: describing the context, identifying thinking traps, and writing a reframed thought. For thinking trap identification, a fine-tuned language model ranked 13 predefined traps and presented likelihood estimates, from which participants could select up to three. For reframing, the system used retrieval-enhanced in-context generation to produce candidate reframes that participants could select, revise, or replace. This interaction structure was preserved using GPT-4.1.
 The two conditions differ along multiple dimensions, including the range of intervention strategies, the use of multimodal interaction elements, and the ability to dynamically construct interaction sequences. \revise{The study therefore evaluates the combined effect of these design choices as opposed to isolating any single component.} Additional technical details about the control can be found in prior works \cite{sharma2023cognitive, sharma2024facilitating}.

\subsubsection{Intervention Outcomes}

To assess outcomes, we measured perceived stress, stress mindset \cite{crum2013rethinking}, user experience (UEQ-8) \cite{hinderks2017design}, and post-activity perceptions. 
Post-activity measures captured aspects like personalization,  reflection of user input, and enjoyment. Table~\ref{tab:measures} lists them all.

We evaluated two primary outcomes: stress reduction ($Stress_{pre} - Stress_{post}$) and user experience (mean UEQ-8 score). We tested the hypotheses that GUIDE would outperform the Control condition using one-sided Welch’s two-sample $t$-tests ($\alpha = 0.05$) with Benjamini–Hochberg correction. Stress mindset improvement and post-activity perceptions were analyzed as exploratory outcomes. We also analyzed open-ended responses about perceived stress impact, helpful components, mismatches, and personalization using thematic analysis~\cite{clarke2017thematic}.

\section{Results}

We present findings from our deployment comparing GUIDE and the control condition on stress reduction, user experience, and related outcomes.

\begin{table}[t!]
\centering
\small
\caption{Summary of outcome measures for participants in the GUIDE and Control conditions.}
\label{tab:outcomes}
\begin{tabular}{@{} p{2.4cm} p{1.3cm} p{1.3cm} p{0.2cm} p{1.3cm} @{}}
\toprule
\textbf{Outcome} & \textbf{GUIDE} & \textbf{Control} & \textbf{p} & \textbf{Cohen's $d$} \\
& (M $\pm$ SD) & (M $\pm$ SD) & & \\
\midrule

\multicolumn{5}{c}{\textbf{Primary outcomes}} \\
\midrule
Stress reduction* & 0.65 $\pm$ 0.7 & 0.35 $\pm$ 0.8 & .02 & 0.39 \\
\midrule
User experience* & 0.49 $\pm$ 0.6 & 0.33 $\pm$ 0.6 & .04 & 0.27 \\
\midrule

\multicolumn{5}{c}{\textbf{Exploratory outcomes}} \\
\midrule
\makecell[l]{Stress mindset \\ improvement} & 1.44 $\pm$ 3.1 & 0.75 $\pm$ 3.5 & .09 & 0.21 \\
\midrule
\makecell[l]{Perceived \\personalization} & 3.39 $\pm$ 1.1 & 3.40 $\pm$ 1.1 & .60 & -0.01 \\
\midrule
\makecell[l]{Perceived system \\ understanding} & 3.44 $\pm$ 1.0 & 3.31 $\pm$ 1.0 & .20 & 0.13 \\
\midrule
\makecell[l]{Perceived reflection \\of user input*} & 3.70 $\pm$ 0.9 & 3.43 $\pm$ 1.0 & .04 & 0.30 \\
\midrule

\makecell[l]{Intent to reuse \\ activity*}& 3.26 $\pm$ 1.0 & 2.90 $\pm$ 1.2 & .03 & 0.33 \\
\midrule
Intent to recommend activity & 3.22 $\pm$ 1.0 & 3.00 $\pm$ 1.2 & .09 & 0.20 \\
\midrule
\makecell[l]{Activity length \\appropriateness} & 3.43 $\pm$ 1.0 & 3.49 $\pm$ 1.1 & .65 & -0.05 \\
\midrule
Activity enjoyment* & 3.44 $\pm$ 1.0 & 3.16 $\pm$ 1.0 & .04 & 0.28 \\
\bottomrule

\end{tabular}
~\\~\\
\footnotesize{* $p < .05$, ** $p < .01$, *** $p < .001$}
\end{table}


\subsection{GUIDE Reduces Stress and Improves User Experience Compared to Control}
\subsubsection{Stress Reduction}

Table \ref{tab:outcomes} presents summary statistics for all outcome measures. Participants in the GUIDE condition showed greater reductions in stress compared to the Control condition ($M_G = 0.65 \pm 0.7$, $M_C = 0.35 \pm 0.8$, $p = .02$, $d = 0.39$), supporting our first primary hypothesis. Improvements in stress mindset were directionally positive but did not reach statistical significance ($p = .09$). The two conditions were comparable at baseline. There were no differences in pre-intervention situation stress (GUIDE: $M = 3.89 \pm 0.8$, Control: $M = 3.84 \pm 0.9$, $p = .71$) or overall perceived stress (PSS; GUIDE: $M = 21.29 \pm 6.0$, Control: $M = 20.59 \pm 6.3$, $p = .39$).

We fit a regression model predicting post-intervention stress while controlling for pre-intervention stress, PSS, age, race, gender, and pre-intervention stress mindset. Participants in the GUIDE condition reported lower post-intervention stress than those in the Control condition ($\beta = -0.28$, $p = .006$). Pre-intervention stress remained a strong predictor of post-intervention stress ($\beta = 0.59$, $p < .001$), while PSS and pre-intervention stress mindset were not significant. These results indicate that the advantage of GUIDE in reducing stress persists after accounting for pre-intervention stress and related individual differences.

We conducted an exploratory analysis comparing GUIDE participants who received cognitive restructuring alone ($n = 17$; see Appendix~\ref{app:cbt-mapping} for mapping to broad CBT categories) with the Control condition, which also used cognitive restructuring. GUIDE showed greater stress reduction ($M_{GCR} = 0.88 \pm 0.60$) than Control ($M_C = 0.35 \pm 0.83$; $p = .002$, $d = 0.74$). \revise{Since GUIDE assigned this intervention based on user context, the subgroup was not formed through randomization, and we interpret this comparison descriptively.} We additionally tested whether differences could be explained by variation in time spent; controlling for time did not change the results ($p = .005$), and time was not associated with outcomes ($p = .72$). 


\subsubsection{User Experience}

Participants in the GUIDE condition reported higher overall user experience scores (UEQ-8, scaled from $-2$ to $+2$) compared to the Control condition ($M_G = 0.49 \pm 0.6$, $M_C = 0.33 \pm 0.6$, $p = .04$, $d = 0.27$), supporting our second primary hypothesis.
This pattern was also reflected in specific aspects of the experience (as shown in Table~\ref{tab:outcomes}). Participants in GUIDE reported higher agreement that the activity reflected what they shared, greater enjoyment of the activity, and a higher willingness to reuse the activity. In contrast, for other aspects of the experience, such as perceived personalization, perceived system understanding, and activity length appropriateness, we did not find evidence that GUIDE was higher than the Control condition (Table~\ref{tab:outcomes}).  Additionally, using the same regression specification as for stress outcomes, \revise{the effect of condition on user experience was weaker after adjustment and the overall model was not significant ($p = .17$)}.





%


\subsection{GUIDE Enabled Diverse Forms of Intervention Experiences}

\begin{figure*}[t]
\centering

\begin{subfigure}[b]{0.42\linewidth}
    \centering
    \includegraphics[width=\linewidth]{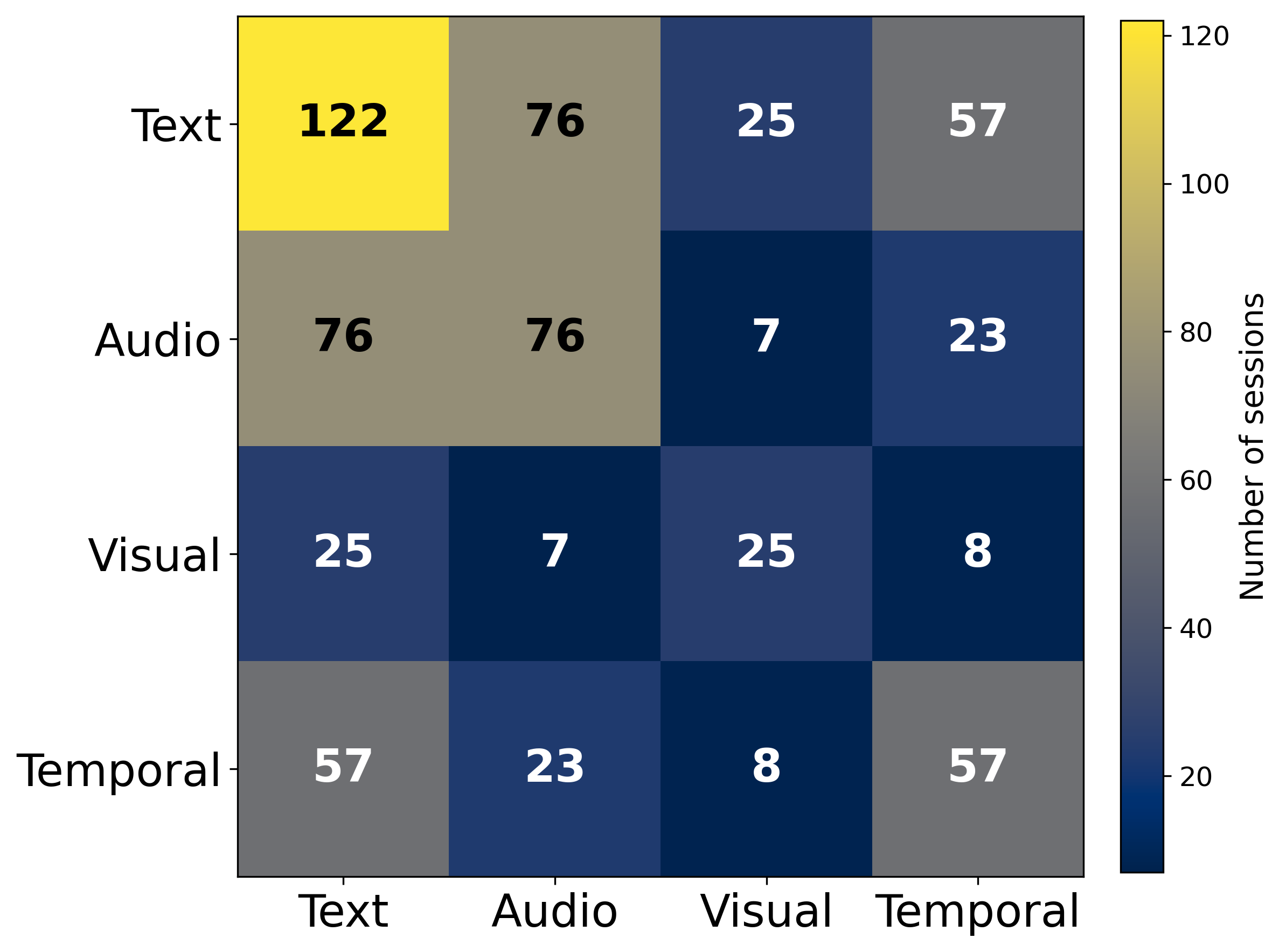}
    \vspace{-0.4em}
    {\small (a)}
\end{subfigure}
\hfill
\begin{subfigure}[b]{0.52\linewidth}
    \centering
    \includegraphics[width=\linewidth]{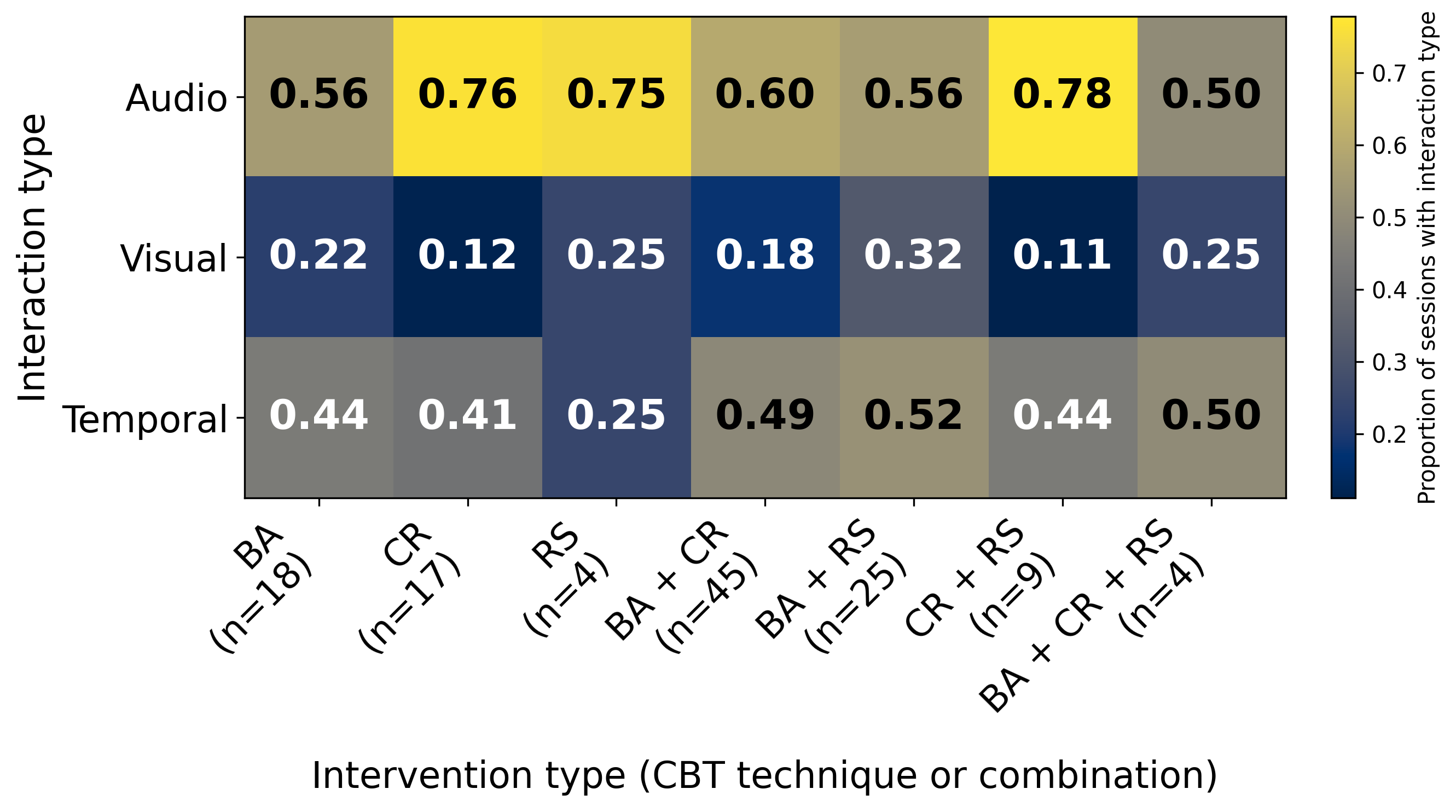}
    \vspace{-0.4em}
    {\small (b)}
\end{subfigure}

\caption{
(a) Co-occurrence of interaction types across generated intervention experiences (diagonal = total usage; off-diagonal = pairwise co-occurrence). 
(b) Interaction type usage across intervention combinations. Values represent the proportion of sessions within each combination that include the interaction type; CR = Cognitive restructuring, BA = Behavioral activation, RS = Regulation strategies.
}
\label{fig:interaction_mapping_combined}
\end{figure*}
We examined the range and structure of the interaction experiences generated by the system. As shown in Figure \ref{fig:interaction_mapping_combined}(a), GUIDE produced experiences that combined multiple interaction types, including text-based interactions (122, 100\%), audio (76, 62.3\%), temporal elements (57, 46.7\%), and visual components (25, 20.5\%). These interaction types were often used together, with most sessions involving multiple forms of interaction, most commonly two (88, 72.1\%).
GUIDE also composed interventions from multiple CBT techniques, most commonly behavioral activation (92, 75.4\%) and cognitive restructuring (75, 61.5\%), with regulation strategies less frequent (42, 34.4\%). These techniques were often combined, particularly behavioral activation and cognitive restructuring (49, 40.2\%), and most sessions involved multiple techniques (79, 64.8\%).

We then examined how interaction types were used to realize intervention techniques across sessions, as shown in Figure \ref{fig:interaction_mapping_combined}(b). Each interaction type was used across multiple techniques and their combinations rather than being tied to a specific technique. For example, audio-based interaction appeared across behavioral activation (56\%), cognitive restructuring (76\%), and their combinations (60–78\%), while temporal elements were similarly used across techniques and combinations (41–52\%, except in regulation strategies). Visual components appeared less frequently but still across different techniques and combinations (11–32\%). This pattern indicates that GUIDE composes intervention experiences by flexibly reusing interaction types across different techniques and combinations, instead of assigning a fixed interaction structure to each intervention technique.

Participants’ qualitative comments reflected this diversity in support. GUIDE enabled multiple forms of support, including reflection, perspective shifting, and concrete action. In contrast, Control participants often expressed a need for more varied and interactive support beyond cognitive restructuring alone. These comments align with GUIDE’s design, which incorporates both thought-focused and action-focused activities, as reflected in participants’ experiences below. 

\subsubsection{Supporting Reflection and Shifting Perspectives Through Guided and Multimodal Interaction}

Participants expressed that GUIDE helped them see their situation more clearly and think about it in a more organized way. Many accounts reflected shifts in perspective, where participants described viewing the problem more broadly, recognizing patterns in their thinking, or feeling a greater sense of control. They described being able to step back and examine it more deliberately. GP57 described this process:

\italquote{I can see a much clearer picture of my overall situation, it feels smaller and more manageable, the thoughts about the situation are not something to be avoided anymore.}

Participants often connected these shifts to how the interaction supported reflection through its multimodal and interactive design. Features such as the ability to record oneself allowed participants to articulate and engage with their thoughts more actively. For example, GP55 noted:

\italquote{Recording a voice note helped me verbalize my thoughts, which also helped me make them clear to myself… this can make the core issue more evident, and easily approachable.}

Participants also described how audio components  supported reflection from different perspectives. Listening to AI guidance or summaries made it easier to step back and reconsider their situation. As GP81 described, this involved \textit{``putting the problem into perspective.''}  Others noted that the audio elements reflected their input closely, such that \textit{``I could see that it [AI Audio] used context and discussed the situations that I described, even minute little details.''} (GP3). Hearing their situation articulated back to them enabled a more detached, third-person view, which made it easier to evaluate their thoughts and consider alternative interpretations.


\subsubsection{Supporting Action and Progress Through Small, Immediate Steps}

Participants described how GUIDE supported them in moving from reflection to taking small, concrete actions. Rather than requiring large changes, the interaction often encouraged manageable steps that felt easier to begin and complete. Several participants noted that focusing on small actions or even brief engagement helped reduce a sense of overwhelm and created a feeling of progress. GP51 described this clearly:

\italquote{Focusing on small steps instead of the whole process helped me feel more in control. Saying the kind line out loud made me notice that progress is real, even if it feels slow.}


Participants also connected this shift to specific UX primitives. Short, time-bound activities that used timers were frequently mentioned as helping them initiate tasks and maintain focus. For example, GP3 noted that ``\textit{getting the short timer to at least get started… did help reduce my stress.}'' These prompts also extended to simple, concrete suggestions such as reaching out to others. GP43 and GP58 appreciated being encouraged to draft short messages that could be used to contact friends or family members.

In addition, list-based activities supported action by helping participants organize their thoughts and identify concrete next steps. These elements guided participants toward what they could do next in their own context. For example, GP66 noted, ``\textit{It helped organize my thought and next steps for today}'', while GP50 described how the activity focused on ``\textit{what I will apply to my next study session}''. These structured prompts helped translate reflection into actionable planning within the interaction.

We also examined whether interaction types (audio, visual, or temporal) were associated with stress reduction within GUIDE. Controlling for pre-stress, no interaction type was associated with outcomes within GUIDE ($p > .30$ for all), \revise{though this exploratory analysis may be underpowered}.

\subsection{GUIDE Generated Diverse but Structured Interaction Sequences}


We analyzed the diversity and organization of interaction sequences for interaction primitives ($\tau$) using information-theoretic and sequence-based measures; formal definitions of these metrics are provided in Appendix \ref{app:diversity-structure}.
GUIDE exhibits high diversity in modular primitive usage, with a normalized entropy of 0.87 (on a 0--1 scale, where higher values indicate a more uniform distribution of available interaction primitives). \revise{Across 122 sessions, GUIDE produced 76 unique interaction sequences.} GUIDE also produced varied sequences with an average sequence similarity of 0.40 (on a 0–1 scale), indicating that sessions share common building blocks but differ in how these primitives are arranged. In contrast, the control condition follows an identical sequence across all sessions (sequence similarity = 1.00).

To assess whether GUIDE's variation reflects meaningful structure and not arbitrary composition, we compare its sequences against shuffled baselines that preserve the same modules within each session but randomly permute their order. This isolates the role of ordering: any differences between observed and shuffled sequences reflect non-random organization rather than differences in content. Under this comparison, GUIDE exhibits higher sequence similarity than shuffled sequences (0.40 vs. 0.36, $p < .01$), and lower transition entropy (0.87 vs. 0.90, $p < .01$). Here, transition entropy measures how predictable the next interaction primitive is given the current one; lower values indicate more predictable transitions between modules. Together, these results show that GUIDE’s interaction flows are more structured and less arbitrary than would be expected if primitives were arranged at random.

We further examine local interaction structure using $n$-grams, which capture short contiguous patterns of modules. 
The most frequent bigram (\texttt{Audio Message} $\rightarrow$ \texttt{Text Input}) occurs in 10.7\% of transitions in GUIDE, compared to 5.8\% in shuffled sequences, while the most frequent trigram (\texttt{List Entry Input} $\rightarrow$ \texttt{Audio Message} $\rightarrow$ \texttt{Text Input}) occurs in 5.4\% of cases compared to 1.6\%. These substantial increases over the shuffled baseline indicate that GUIDE repeatedly assembles certain combinations of interaction primitives, forming recurring structural motifs rather than arbitrary sequences.

These findings show that GUIDE generates diverse interaction sequences with non-random organization. In contrast to the fixed structure of the control condition, GUIDE supports flexible module composition and produces recurring local patterns in combining interaction primitives.

\subsection{Perceived Personalization Varied Across the Interaction Sequence}


Although GUIDE adapted intervention content based on user input, we did not find evidence that perceived personalization was higher in GUIDE than in the Control condition (Table~\ref{tab:outcomes}; $M_G = 3.39 \pm 1.1$, $M_C = 3.40 \pm 1.1$, $p = .60$). 
This suggests that participants may have perceived both conditions as similarly personalized, and may not have clearly distinguished between personalization in content and personalization in how the interaction was structured. At the same time, qualitative responses indicate that personalization was present but not always consistently experienced across GUIDE.

Participants described instances of how GUIDE translated their situation into concrete, actionable steps by operationalizing details from their input into what they were asked to do. In these cases, the activity was directly shaped by what participants were currently dealing with, rather than only reflecting it in language. For instance, GP1 said,

\italquote{[The system] got my input about my situation and the courses I was taking and it used that to specifically prompt me in thinking ahead about my studies and career.}

This grounding of context often extended into situation-specific interventions where user-provided details shaped the structure of the activity itself. GP76 described how the system used the fact that they were already listening to music to design an activity around it. Similarly, GP110 noted that the activity prompted them to spend time on a specific topic, such as the ``softmax function'' (a machine learning concept in their syllabus), before reflecting on their progress.

At the same time, participants described that this sense of personalization did not always extend consistently across their entire activity. For instance, GP16 described receiving a breathing activity before writing a plan; while they appreciated the setup, they did not see it as tailored to their situation. Similarly, GP31, who received a reflection activity following a timer-based prompt to make progress, noted that the timer component did not feel personalized to them, even though the earlier reflection steps felt more relevant. These responses reflect how participants experienced variation within a single activity, where some parts felt closely tied to their situation while others felt less personalized.





\section{Discussion}




\revise{We examined generative experience, an approach to DMH support in which the intervention experience is constructed at runtime. We evaluated GUIDE as an intervention system~\cite{slovak2024hci},} and compared to an LLM-based cognitive restructuring control, it produced greater stress reduction and better user experience. In this section, we discuss the implications of these findings and outline key limitations and directions for future work.

\subsection{Expanding Support Through Diverse Intervention Pathways}
GUIDE operated over a broad space of CBT-informed approaches and constructed interaction forms aligned with their intended function. Voice-based components supported restructuring by helping participants hear their situation from a different perspective \cite{vowels2025evaluating}, while timed activities and structured inputs supported behavioral activation by helping users initiate and sustain action \cite{wennberg2018effectiveness}. Participants’ accounts reflected these proximal effects, describing how the activity helped them clarify their thoughts, gain perspective, and identify concrete next steps. At the same time, interaction types were not tied to specific techniques, but appeared across different intervention pathways in varying combinations, suggesting that GUIDE flexibly reused interaction primitives to support similar psychological processes in different ways.

More broadly, these findings highlight the value of generative experience as a design concept. In clinical and other helping contexts, support is rarely delivered as a fixed technique in a fixed format, but is adapted to the person and situation \cite{bhattacharjee2023investigating, stiles1998responsiveness, chorpita2005modularity}. GUIDE moves toward this model by treating intervention support as a compositional problem, translating intervention specifications into sequences of interaction primitives \cite{cao2025generative, chen2025generative}. Our exploratory analyses are consistent with gains not being attributable to intervention type, specific modalities, or time spent alone, \revise{though they cannot isolate mechanism}. Taken together with participants’ reports, this may point to the possibility that benefits may have arisen from how intervention and interaction were combined within the activity, in ways that better supported reflection, perspective taking, and action planning. This suggests that generative systems may be useful when they vary how support unfolds while still preserving enough structure to guide users through the intended psychological process.

\subsection{Balancing Personalization Across Interaction Sequences}


As generative interfaces open a combinatorial design space of UX elements, the challenge shifts toward ensuring that these elements work together as a coherent whole \cite{persson2025design, wangmi2015framework, slovak2024hci}. Participants’ comments suggest that personalization is experienced across the interaction rather than within isolated components, with moments where the activity felt closely tied to their situation and others where it felt more generic. This highlights the importance of continuity across steps, where each part builds on prior context, and breaks in this continuity can reduce the overall sense of personalization \cite{slovak2024hci}. In GUIDE, the UX sequence was generated after an initial context elicitation phase, enabling tailoring to the user’s situation, though future systems could adopt more stepwise approaches where later steps build on earlier responses.

At the same time, uniform personalization across the full sequence may not be necessary \cite{bhattacharjee2025perfectly}. Some activities (e.g., breathing for GP16) may feel less specific at the level of a single step, yet still contribute to overall outcomes when embedded within a broader experience shaped by user context. This may suggest that not every component  needs to be highly tailored, as long as the overall interaction remains responsive to the user’s situation. Future work should examine how personalized and general components are combined within an experience, and when broader activities are sufficient versus when stronger contextual tailoring is needed.

\subsection{Considerations for Broader Applications}
While our study focuses on an SSI, extending this approach to longitudinal use and other mental health contexts introduces additional challenges. The current design relies on guided elicitation to capture users’ situations, but repeatedly asking for detailed context may become burdensome over time \cite{bhattacharjee2024exploring}. In longer-term use, systems may need to maintain an evolving representation of user context across sessions, combining lightweight input with temporal patterns, device usage, or other passive data sources \cite{huckins2020mental, xu2023globem, mohr2017personal}. This would allow interventions to remain responsive without requiring full re-articulation each time. Extending beyond stress to conditions such as anxiety or depression may also require adapting the intervention strategies and interaction structures that are generated.

Beyond DMH, generating experiences at runtime may apply to other domains where support is delivered through structured interaction. In areas such as education, coaching, or behavior change, systems often rely on fixed formats even when content is personalized \cite{kazemitabaar2024codeaid, wu2024mindshift, jorke2025bloom}. A generative experience approach could enable systems to construct activities that fit different user needs, for example by helping one student trace a bug in their own code step by step, while helping another review the specific kinds of errors they are most likely to make in an education setting. However, similar challenges would arise, including maintaining coherence across interactions and ensuring alignment with domain-specific principles. These considerations suggest that generative experience may generalize beyond DMH, while requiring adaptation to the goals and constraints of each domain.

\subsection{Limitations and Future Work}

This work has several limitations. First, we cannot cleanly disentangle the effects of generative experience from intervention diversity. GUIDE draws from a broad set of CBT-informed strategies, whereas the control focuses on a single cognitive restructuring workflow, so improvements may reflect differences in what participants were asked to do as well as how the experience was generated. \revise{Because we evaluated GUIDE as an integrated intervention system, component-level ablations with real users remain an important direction,} for example, comparing multiple interaction realizations of the same intervention. \revise{The number of candidates at each stage ($n = 3$) was fixed and not empirically optimized; future work should systematically vary $n$ to examine tradeoffs among diversity, latency, and output quality.}

Second, our evaluation context is limited. \revise{The study was conducted with students from a single course in a single-session setting with a novel interaction format, which may limit generalizability and introduce novelty effects \cite{poppenk2010revisiting}. Future work should examine this approach across more diverse populations and under repeated use to assess longer-term engagement and outcomes.}

Third, while GUIDE incorporates a range of interaction primitives, other modalities (e.g., haptics \cite{pacheco2024haptic}, embodied agents \cite{provoost2017embodied}) may further expand how support can be provided. In addition, GUIDE relies on a particular AI generation pipeline and language model, and system behavior may depend on model capabilities. Future work should explore broader design spaces and examine robustness across models. \revise{Relatedly, while participants' accounts and our CBT mapping (Appendix \ref{app:cbt-mapping}) seem to be consistent with enactment of the intended mechanisms, we did not conduct a formal clinical fidelity evaluation of the generated interventions; formal fidelity rating remains an important direction for future work.}

\begin{acks}
This work was supported by funding from the Stanford King Center on Global Development. We thank members of the AI4HI Lab and the SALT Lab for their thoughtful feedback, discussions, and support throughout the project, with special thanks to Ryan Louie. We are also grateful to members of the CREATE Center at Stanford who tested early versions of the system and shared valuable feedback. Finally, we thank Blanca Tezanos and Irina Lechtchinskaia for their assistance with financial and administrative support.
\end{acks}

\bibliographystyle{ACM-Reference-Format}
\bibliography{main}
\newpage
\appendix
\onecolumn
\section{Expert Consultation Procedure}
\label{app:expert}

We recruited six experts through personal networks and an open call on Upwork. Experts met at least one of the following criteria: (1) formal training or published work in clinical psychology, counseling psychology, behavioral science, or health-related HCI, or (2) at least three years of experience working with psychological interventions or stress management.

The mean age was 37.8$\pm$9.2 years, and participants had an average of 11.5$\pm$8.2 years of experience working with psychological interventions. The sample included individuals from multiple racial backgrounds (4 White, 1 African American, and 1 mixed race). Three participants held Master’s degrees, and three held Doctoral degrees. The group included four men and two women.

We conducted semi-structured consultation sessions spanning an hour through Zoom videoconferencing platform. Each expert was first introduced to the system and the overall goal of generating contextualized stress management support. They then engaged directly with the tool by interacting with it across multiple stress scenarios while sharing their screen, allowing the interviewer to observe their process. During and after these interactions, experts provided feedback on the system’s context elicitation process, the quality and appropriateness of generated interventions, and the overall user experience. The interview questions focused on evaluating how well the generated support fit the user’s situation, whether the steps and framing aligned with the provided context, and the extent to which the interventions reflected established psychological principles. Experts were also asked to assess potential risks or mismatches, critique and refine the rubric used for generation and evaluation, and suggest improvements to both the intervention design and interaction flow. Each expert received \$50 USD for their participation.

\section{Ablation Study}
\label{app:ablation}

Before conducting the main user study, we performed an ablation study to examine the role of rubric guidance in intervention  and UX generation. We compared four conditions: Intervention rubric + UX rubric, No intervention rubric + UX rubric, Intervention rubric + No UX rubric, and No intervention rubric + No UX rubric. When rubrics are present, the system generates multiple candidate interventions or UX structures and selects among them using the corresponding rubrics. When rubrics are absent, the system directly prompts an LLM to produce a single intervention or UX structure without rubric based selection.

The study was conducted using 15 simulated user contexts representing common stress situations experienced by young adults, including academic or career pressure, relationship difficulties, major life transitions, and uncertainty about the future. Each simulated interaction involved two roles. The System Agent executed the intervention pipeline exactly as it would during a real user interaction, generating intervention content and assembling the UX structure. A User Simulator represented the participant side of the interaction. The simulator was assigned a stress persona corresponding to one of the predefined contexts and completed the same chat based context provision step used in the actual system (i.e., elicitation of stress context). The System Agent then produced the intervention and UX structure, following the same generation process used with human users.

To compare outputs, we used an LLM evaluator that assessed the four condition outputs for each context on two outcomes: predicted stress change and predicted UX score. Predicted stress change reflects how much the generated intervention is expected to reduce the user’s stress, while predicted UX score reflects the clarity, structure, and usability of the interaction flow. For each context, the evaluator ranked the four outputs from 1 to 4 for each outcome, where rank 1 indicates the best performing condition. 

\begin{figure*}[!t]
    \centering
\includegraphics[width=0.90\linewidth]{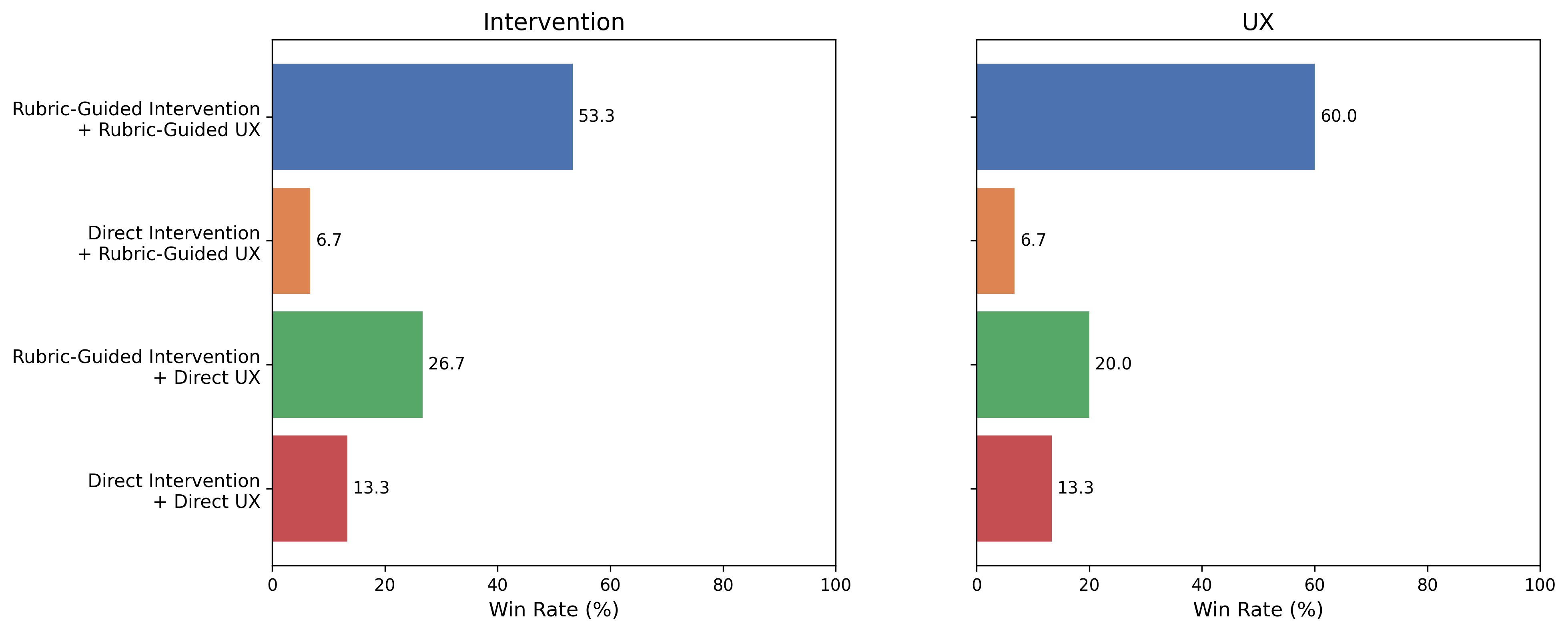}
    \caption{Ablation study results across four conditions defined by the presence or absence of rubric guidance in intervention and UX generation}
    \label{fig:top_1}
\end{figure*}

Figure \ref{fig:top_1} shows the percentage of contexts in which each condition was selected as best. For predicted stress change, the Intervention rubric + UX rubric condition was ranked first in 53.3\% of contexts, followed by Intervention rubric + No UX rubric (26.7\%), No intervention rubric + No UX rubric (13.3\%), and No intervention rubric + UX rubric (6.7\%). A similar pattern is observed for predicted UX score, where the Intervention rubric + UX rubric condition is ranked first in 60.0\% of contexts.

\revise{We note that this ablation served as a configuration check during system development: findings are based on simulated user contexts and should be interpreted as preliminary evidence that rubric-guided generation may improve system outputs, not as a component-level evaluation with real users.} We therefore use the full pipeline in the subsequent user study and compare the complete system against an established single session activity in a real user setting.

\section{Interaction Primitives}
\label{app:primitives}

\begin{figure*}[!t]
\centering
\captionsetup[subfigure]{font=scriptsize, justification=centering}

\begin{subfigure}[c]{0.48\textwidth}
\centering
\begin{minipage}[c]{\linewidth}
\centering
\includegraphics[height=4.8cm, width=0.94\linewidth]{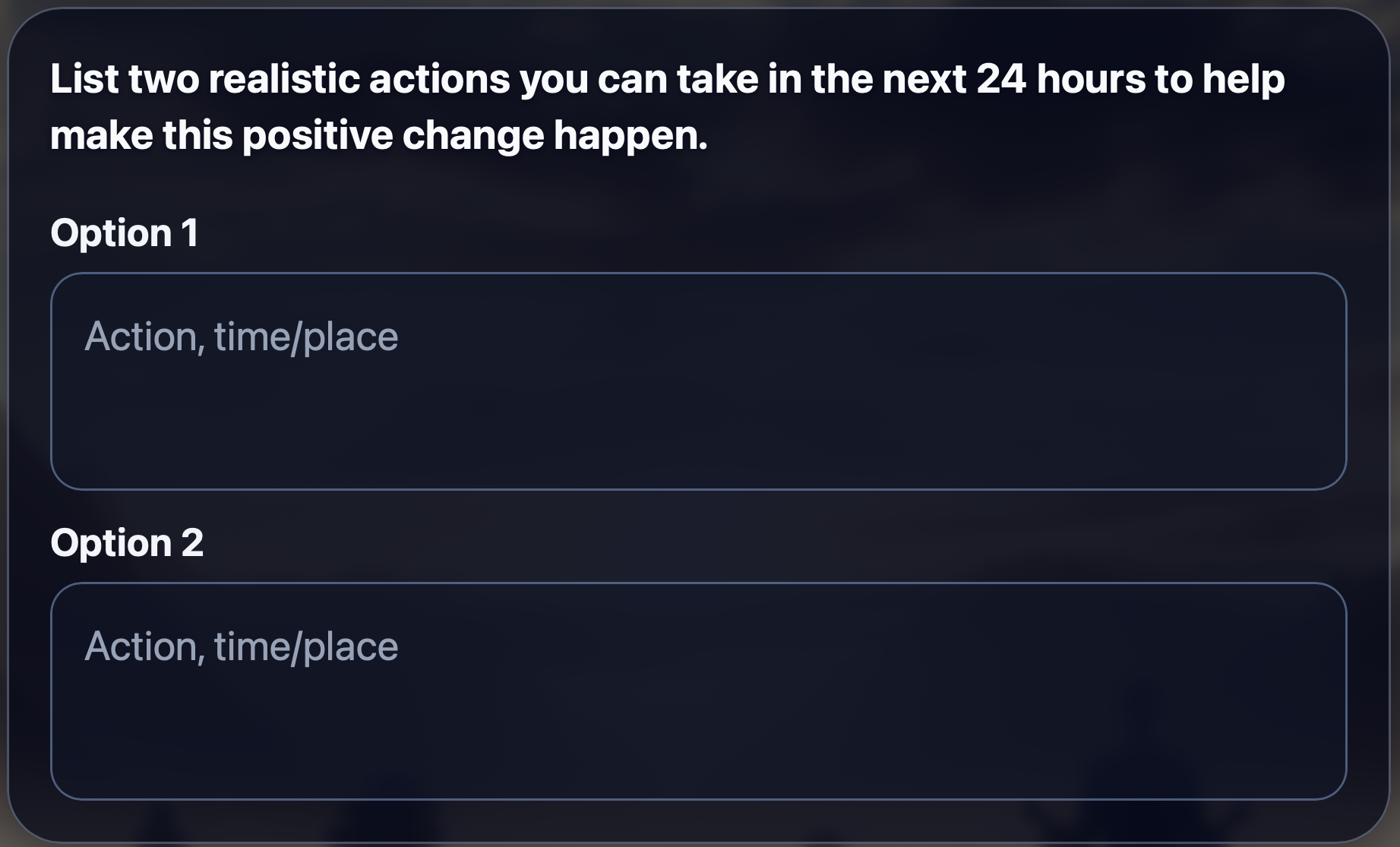}
\end{minipage}
\caption{Text-based Interaction}
\end{subfigure}
\hfill
\begin{subfigure}[c]{0.48\textwidth}
\centering
\begin{minipage}[c]{\linewidth}
\centering
\includegraphics[height=4.8cm, width=0.98\linewidth]
{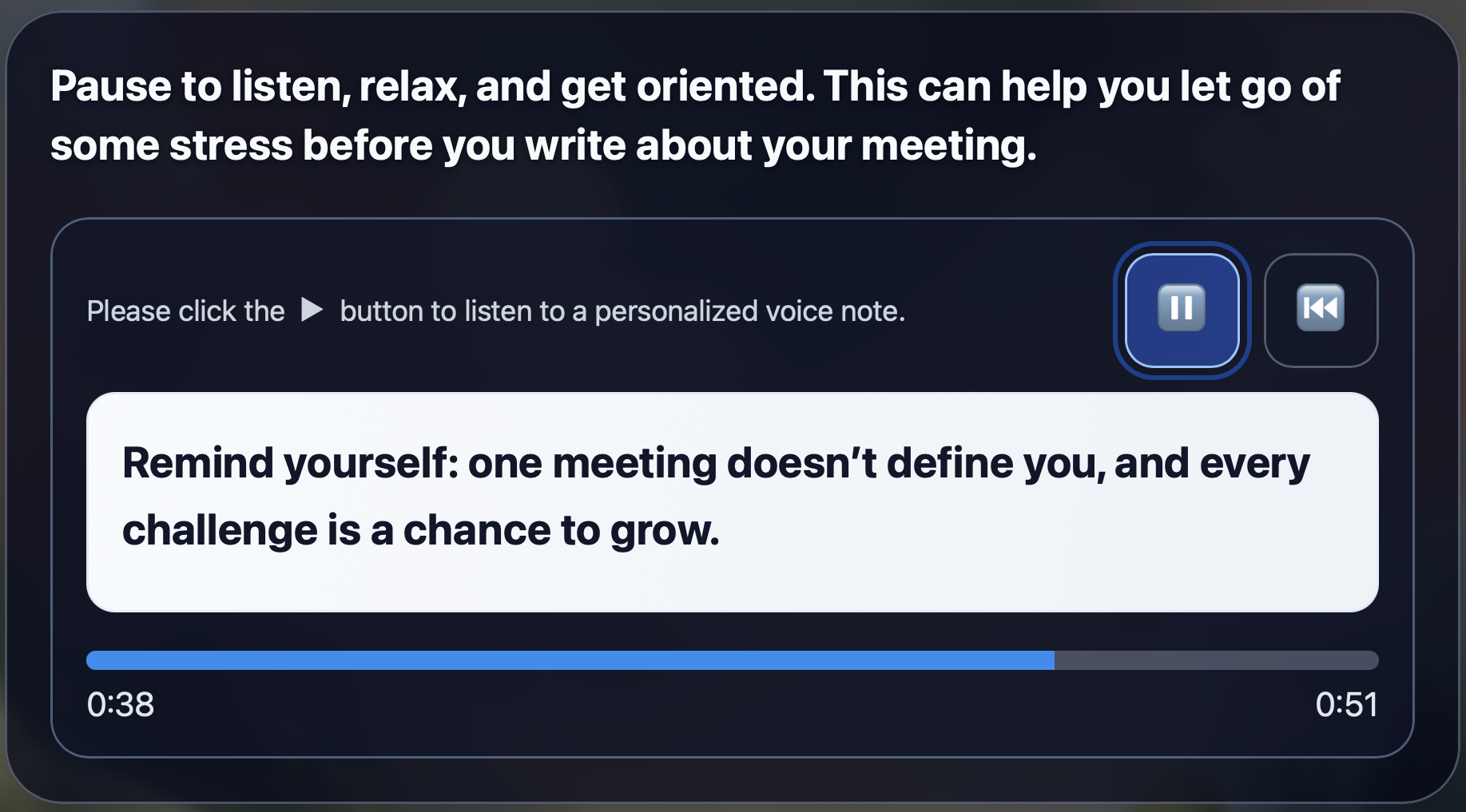}
\end{minipage}
\caption{Audio-based Interaction}
\end{subfigure}

\vspace{0.6em}

\begin{subfigure}[c]{0.48\textwidth}
\centering
\begin{minipage}[c]{\linewidth}
\centering
\includegraphics[width=0.94\linewidth]{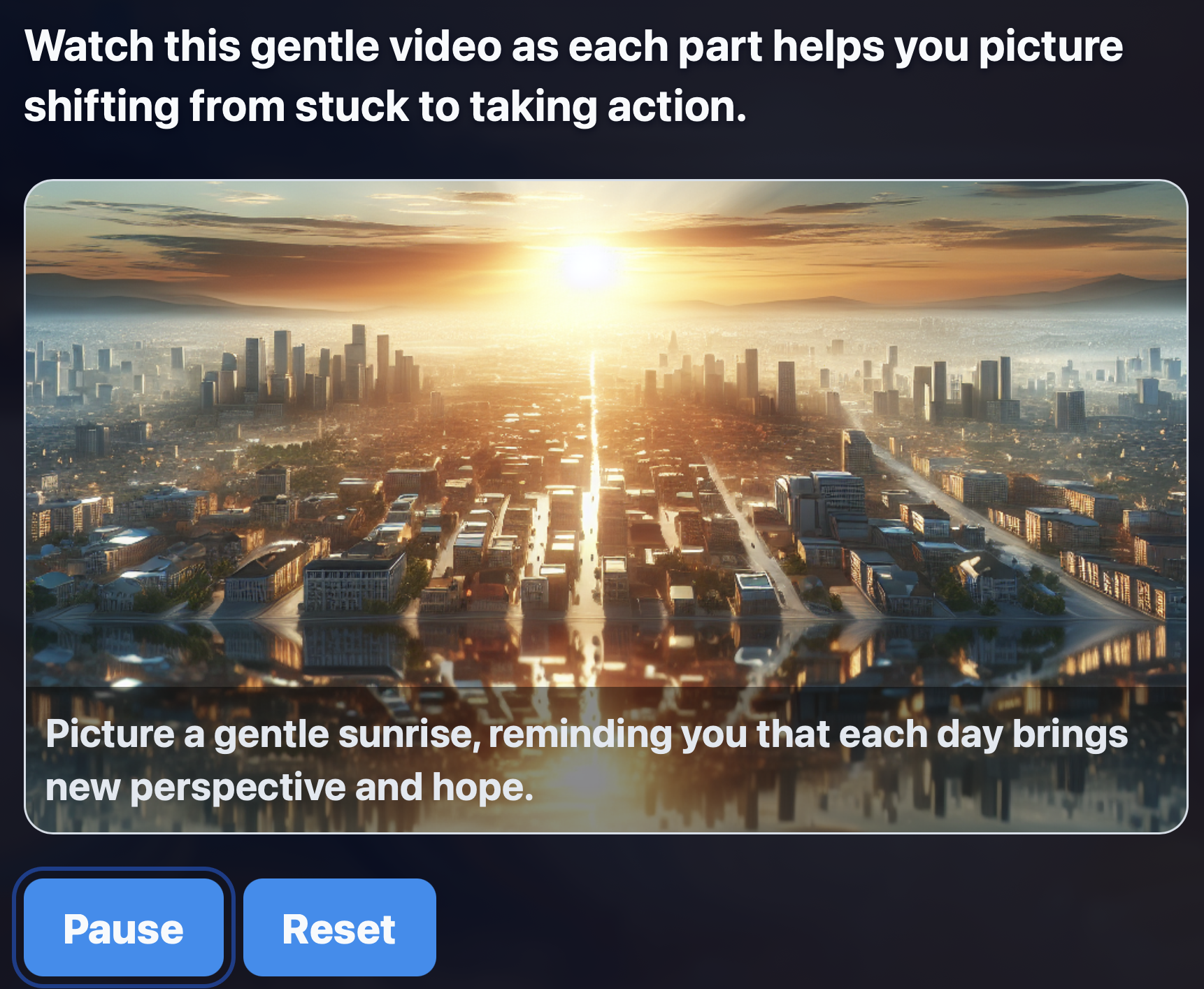}
\end{minipage}
\caption{Visual Interaction}
\end{subfigure}
\hfill
\begin{subfigure}[c]{0.48\textwidth}
\centering
\begin{minipage}[c]{\linewidth}
\centering
\includegraphics[ width=0.98\linewidth]{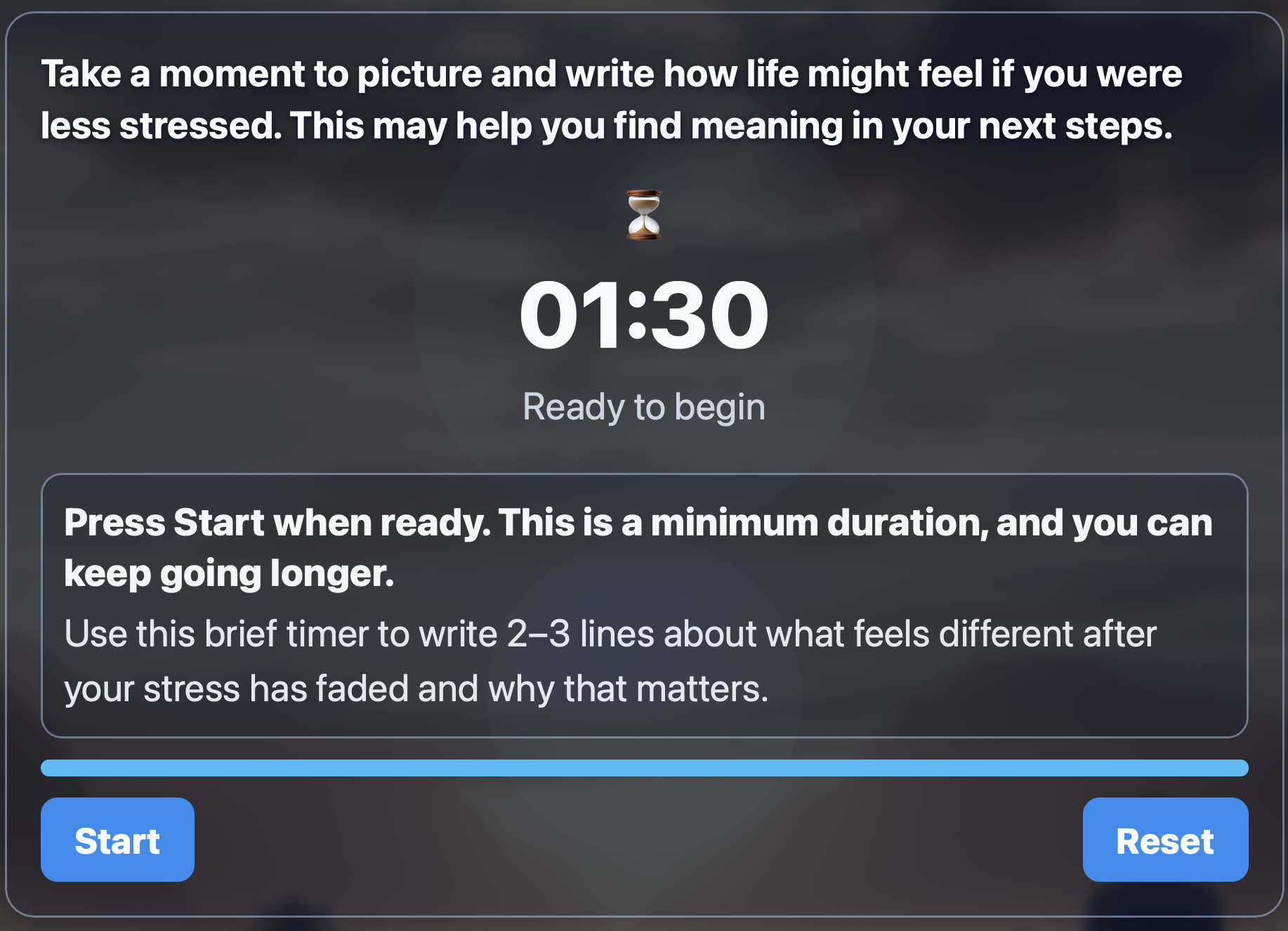}
\end{minipage}
\caption{Temporal Interaction}
\end{subfigure}

\caption{Illustrative examples of interaction types used to construct intervention experiences. In practice, each type can take many forms and be combined in different ways depending on the intervention and user context (See Figures~\ref{fig:exam_example}, ~\ref{fig:meeting_example}, and ~\ref{fig:familymember_example}).}
\label{fig:ux_palette}
\end{figure*}

We detail the interaction primitives, along with their parameters and associated interaction types in Table~\ref{tab:ux_palette}, and illustrate example interfaces in Figure~\ref{fig:ux_palette}.

\begin{table}[t]
\centering
\small
\caption{Overview of interaction primitives, their parameters, and interaction types.}
\label{tab:ux_palette}
\setlength{\tabcolsep}{4pt}
\begin{tabular}{@{} p{2.9cm} p{6.8cm} p{2.3cm} @{}}

\toprule
\textbf{Primitive ($\tau$)} & \textbf{Parameters ($\theta$)} & \textbf{Interaction Type} \\
\midrule

Choice Input &
prompt question; response options; multiple selection setting; intervention purpose &
Text \\

Text Input &
prompt question; response hint; intervention purpose &
Text \\

List Entry Input &
list prompt; item labels; item response hints; intervention purpose &
Text \\

Chatbot &
prompt question; system persona; intervention purpose; conversation history &
Text/Audio \\

Audio Message &
audio script; delivery tone; voice pitch; speaking rate; intervention purpose; guidance rationale &
Text/Audio \\

\makecell[l]{Guided \\ Sequence} &
\makecell[l]{timed cue steps; audio cue script;\\ intervention purpose} &
\makecell[l]{Text/Audio/\\ Temporal} \\

Voice Input &
recording prompt; intervention purpose &
Text/Audio \\

Image Upload &
capture prompt; allowed image sources; intervention purpose &
Text/Visual \\

Image Display &
image description prompt; intervention purpose &
Text/Visual \\

\makecell[l]{Visual Card \\ Pair} &
\makecell[l]{frame titles; frame text; frame\\ image prompts; intervention \\purpose} &
Text/Visual \\

Video Clip &
scene prompts; narration script; intervention purpose &
Text/Visual/Audio \\

Timer &
duration; timer text; completion action; reflection prompt; reflection response hint; intervention purpose &
Text/Temporal \\

\bottomrule
\end{tabular}
\end{table}
\section{Mapping Generated Interventions to CBT}
\label{app:cbt-mapping}

To relate generated interventions to established therapeutic frameworks, we mapped them to Cognitive Behavioral Therapy (CBT) principles. The first author, who has extensive experience designing CBT-based interventions across both AI-mediated and non-AI settings (including 10+ prior works), first met with domain experts (E5 and E6) in separate alignment sessions to establish core CBT constructs and their interpretation in the context of our system, distinct from the prior expert consultations on system design. Together, they collaboratively reviewed and mapped an initial set of 40 intervention outputs to build a shared understanding of how different techniques were expressed in practice. Following this calibration process, the remaining interventions were mapped by the first author in line with these discussions. \revise{We note that this mapping is not a formal clinical fidelity evaluation, which remains future work.}

\section{Metrics for Diversity and Structure}
\label{app:diversity-structure}

We quantify the diversity and structural organization of interaction modules and sequences using information-theoretic and sequence-based measures. Let $\mathcal{S} = \{s_1, \dots, s_N\}$ denote the set of interaction sequences, where each sequence $s_i = (x_1, \dots, x_{L_i})$ is an ordered list of UI primitives drawn from a vocabulary $\mathcal{V}$.

\noindentparagraph{Normalized Entropy of Module Usage.}
To measure diversity in module usage, we compute the normalized Shannon entropy over module frequencies. Let $c(v)$ denote the total count of module $v \in \mathcal{V}$ across all sequences, and define:
\begin{equation}
p(v) = \frac{c(v)}{\sum_{v' \in \mathcal{V}} c(v')}
\end{equation}

The entropy is:
\begin{equation}
H = - \sum_{v \in \mathcal{V}} p(v) \log_2 p(v)
\end{equation}

We normalize entropy to the range $[0,1]$:
\begin{equation}
H_{\text{norm}} = \frac{H}{\log_2 |\mathcal{V}|}
\end{equation}

Higher values indicate more diverse and evenly distributed use of modules.

\noindentparagraph{Sequence Similarity.}
To quantify similarity between two sequences $s_i$ and $s_j$, we use an order-sensitive similarity based on aligned matching subsequences. Let $M(s_i, s_j)$ denote the number of primitives that can be matched between the two sequences while preserving order (as computed by \texttt{SequenceMatcher}). The similarity is defined as:
\begin{equation}
\text{Sim}(s_i, s_j) = \frac{2 M(s_i, s_j)}{|s_i| + |s_j|}
\end{equation}

where $|s_i|$ and $|s_j|$ are the lengths of the sequences. The overall similarity is computed as the average pairwise similarity across all sequence pairs.

\noindentparagraph{Transition Entropy.}
To capture local structural consistency, we compute entropy over transitions between consecutive primitives. Let $T = \{(x_t, x_{t+1})\}$ denote all adjacent pairs observed across sequences. Let $c(a,b)$ be the number of times module $b$ follows module $a$, and define:
\begin{equation}
p(a,b) = \frac{c(a,b)}{\sum_{(a',b')} c(a',b')}
\end{equation}

The transition entropy is:
\begin{equation}
H_{\text{trans}} = - \sum_{(a,b)} p(a,b) \log_2 p(a,b)
\end{equation}

We normalize by the maximum possible entropy:
\begin{equation}
H_{\text{trans}}^{\text{norm}} = \frac{H_{\text{trans}}}{\log_2 |T|}
\end{equation}

Lower values indicate more predictable and consistent transitions.

\noindentparagraph{Shuffled Baseline.}
To assess whether observed structure arises from non-random ordering, we construct a shuffled baseline by randomly permuting each sequence:
\begin{equation}
\tilde{s}_i = \pi(s_i)
\end{equation}
where $\pi$ is a random permutation. This preserves the multiset of primitives and sequence length, but removes ordering structure. All metrics are recomputed on $\{\tilde{s}_i\}$ and compared against observed values.

\noindentparagraph{$n$-gram Analysis.}
We analyze local patterns using $n$-grams. For a sequence $s = (x_1, \dots, x_L)$, an $n$-gram is a contiguous subsequence:
\begin{equation}
g_t^{(n)} = (x_t, x_{t+1}, \dots, x_{t+n-1})
\end{equation}

We compute the frequency:
\begin{equation}
p(g) = \frac{c(g)}{\sum_{g'} c(g')}
\end{equation}

for all observed $n$-grams. We compare $p(g)$ against the corresponding frequency under shuffled sequences to identify patterns that occur more frequently than expected under random ordering.

\newpage
\onecolumn
\section{Additional Details About Judging Process}
\label{app:int-judge-rubric}

\begin{table*}[h!]
\centering
\small
\caption{Intervention judge rubrics used to evaluate the quality of generated interventions.}
\label{tab:int-judge_rubrics}
\begin{tabular}{p{3.5cm} p{7.5cm} p{4cm}}
\toprule
\textbf{Rubric} & \textbf{Description} & \textbf{Rationale} \\
\midrule

Narrative Flow & Whether the reflective and action oriented steps form a clear and continuous experience where each step connects naturally to the next. & Supports coherence in multi-step interventions (prior work \cite{persson2025design}) \\

Small Progress & Whether the sequence ends with a concrete outcome such as a clarified sentence, named emotion, reframe, or small action. & Emphasizes actionable outcomes (expert feedback) \\

Safe Sequencing & Whether the steps remain low intensity, clearly bounded, and easy to pause, avoiding heavy emotional processing. & Reduces risk in brief interventions (prior work \cite{persson2025design}; expert feedback) \\

Explicit Alignment with Psychology Principles & Whether the activity clearly names the psychological principle and demonstrates how the steps enact it. & Improves transparency and learning (prior work \cite{mohr2014behavioral, persson2025design, paredes2014poptherapy}) \\

Specificity & Whether the activity reuses the user's phrases, routines, or constraints so the activity clearly belongs to their situation. & Enhances perceived relevance (prior work \cite{yardley2015person}) \\

Non-retrievability & Whether the activity depends on the user's specific context and would not easily apply to another person. & Avoids generic responses (expert feedback) \\

Everyday Feasibility & Whether the activity can be completed immediately on the user's device within about ten minutes without additional materials. & Supports real-world usability (expert feedback) \\

Understandability & Whether instructions remain simple, clear, and grounded in plain language while reflecting the user's context. & Ensures accessibility and clarity (expert feedback) \\

\bottomrule
\end{tabular}
\end{table*}

\begin{table*}[h!]
\centering
\small
\caption{UX judge rubrics used to evaluate the quality of generated interaction experiences.}
\label{tab:ux_judge_rubrics}
\begin{tabular}{p{3.8cm} p{7.2cm} p{4cm}}
\toprule
\textbf{Rubric} & \textbf{Description} & \textbf{Rationale} \\
\midrule

Intervention-Interface Alignment & Whether the overall structure of the interface reflects the user’s request and presents modules in a clear, stepwise progression. & Ensures alignment between user intent and interaction flow (prior work \cite{nielsen1994usability, hartmann2008towards}) \\

Task Efficiency & Whether the activity can be completed with minimal friction, limited typing, and within the intended time window. & Reduces effort and supports short, feasible interventions (expert feedback) \\

Usability & Whether interactive controls are clear, visible, and actionable, with consistent navigation and feedback. & Improves interaction reliability and ease of use (prior work \cite{nielsen1994usability, hartmann2008towards, brooke1996sus}) \\

Information Clarity & Whether content is structured, concise, and easy to scan, reducing cognitive load. & Supports comprehension and reduces overload (prior work \cite{nielsen1994usability, hartmann2008towards}) \\

Interaction Satisfaction & Whether the experience ends with a clear sense of completion, visual consistency, and smooth transitions. & Reinforces completion and overall experience quality (prior work \cite{nielsen1994usability, hartmann2008towards}; expert feedback) \\

Specificity & Whether the interface incorporates the user’s specific situation through wording, examples, or UI elements. & Enhances contextual relevance and personalization (prior work \cite{yardley2015person}) \\

Understandability & Whether instructions use simple, everyday language aligned with the user’s framing. & Improves accessibility and understanding (expert feedback) \\

\bottomrule
\end{tabular}
\end{table*}


\begin{figure*}
    \centering
    \includegraphics[width=0.95\linewidth]{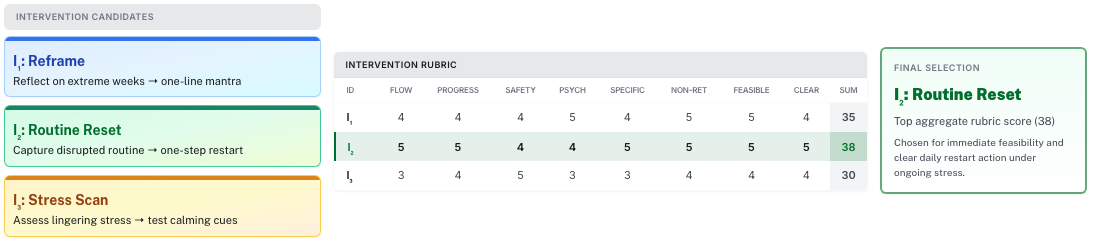}
    \caption{Rubric-guided selection of intervention candidates. Given a user context, multiple candidate interventions are generated and evaluated across rubric criteria, and the highest-scoring intervention is selected. UX structure is also generated in a similar process.}
    \label{fig:judge}
\end{figure*}

\newpage
\section{Regression Results}

\begin{table}[H]
\centering
\small
\caption{Regression predicting post-intervention stress controlling for pre-stress and demographic variables}
\label{tab:reg_stress}
\begin{tabular}{lccccccccc}
\toprule
Outcome & Condition (GUIDE) & Pre-stress & PSS & Stress mindset (pre) & Age & Race (White) & Race (PNA) & Gender (Male) \\
\midrule
Post-stress 
& -0.28** \, (0.10)
& 0.59*** \, (0.08)
& 0.01 \, (0.01)
& -0.01 \, (0.01)
& 0.08* \, (0.04)
& -0.15 \, (0.16)
& -0.28* \, (0.11)
& -0.09 \, (0.13) \\
\bottomrule
\end{tabular}

\vspace{3pt}
\raggedright
\footnotesize{
Entries are coefficient estimates with standard errors in parentheses. Control is the reference group for condition, Asian for race, and Female for gender. PNA = prefer not to answer. * $p < .05$, ** $p < .01$, *** $p < .001$.
}

\end{table}

\begin{table}[H]
\centering
\small
\caption{Regression predicting post-intervention user experience (UEQ) controlling for pre-stress and demographic variables}
\label{tab:reg_ueq}
\begin{tabular}{lccccccccc}
\toprule
Outcome & Condition (GUIDE) & Pre-stress & PSS & Stress mindset (pre) & Age & Race (White) & Race (PNA) & Gender (Male)\\
\midrule
UEQ mean 
& 0.12 \, (0.09)
& 0.08 \, (0.06)
& -0.01 \, (0.01)
& 0.01 \, (0.01)
& 0.01 \, (0.03)
& -0.03 \, (0.15)
& 0.01 \, (0.20)
& -0.05 \, (0.10) \\
\bottomrule
\end{tabular}

\vspace{3pt}
\raggedright
\footnotesize{
Entries are coefficient estimates with standard errors in parentheses. Control is the reference group for condition, Asian for race, and Female for gender. PNA = prefer not to answer. * $p < .05$, ** $p < .01$, *** $p < .001$.
}
\end{table}

\begin{table}[H]
\centering
\small
\caption{Regression predicting post-intervention stress controlling for pre-stress and time spent}
\label{tab:reg_stress_time}
\begin{tabular}{lcccc}
\toprule
Outcome & Condition (GUIDE) & Pre-stress & Log-transformed time \\
\midrule
Post-stress 
& -0.30** (0.10)
& 0.67*** (0.07)
& 0.02 (0.07) \\
\bottomrule
\end{tabular}

\vspace{3pt}
\raggedright
\footnotesize{
Entries are coefficient estimates with standard errors in parentheses. Control is the reference group. * $p < .05$, ** $p < .01$, *** $p < .001$.
}
\end{table}

\begin{table}[H]
\centering
\small
\caption{Regression predicting post-intervention stress within the GUIDE condition using interaction type presence}
\label{tab:reg_modality}
\begin{tabular}{lcccc}
\toprule
Outcome & Pre-stress & Audio & Visual & Temporal \\
\midrule
Post-stress 
& 0.22** (0.08)
& -0.18 (0.18)
& -0.08 (0.20)
& 0.06 (0.15) \\
\bottomrule
\end{tabular}

\vspace{3pt}
\raggedright
\footnotesize{
Entries are coefficient estimates with standard errors in parentheses. Audio, visual, and temporal indicate presence of each interaction type. * $p < .05$, ** $p < .01$, *** $p < .001$.
}
\end{table}
\newpage
\onecolumn

\section{Example Generated Experiences}
\label{app:example}

\begin{figure}[H]
    \centering
    \begin{subfigure}[c]{0.48\linewidth}
        \centering
        \includegraphics[width=\linewidth]{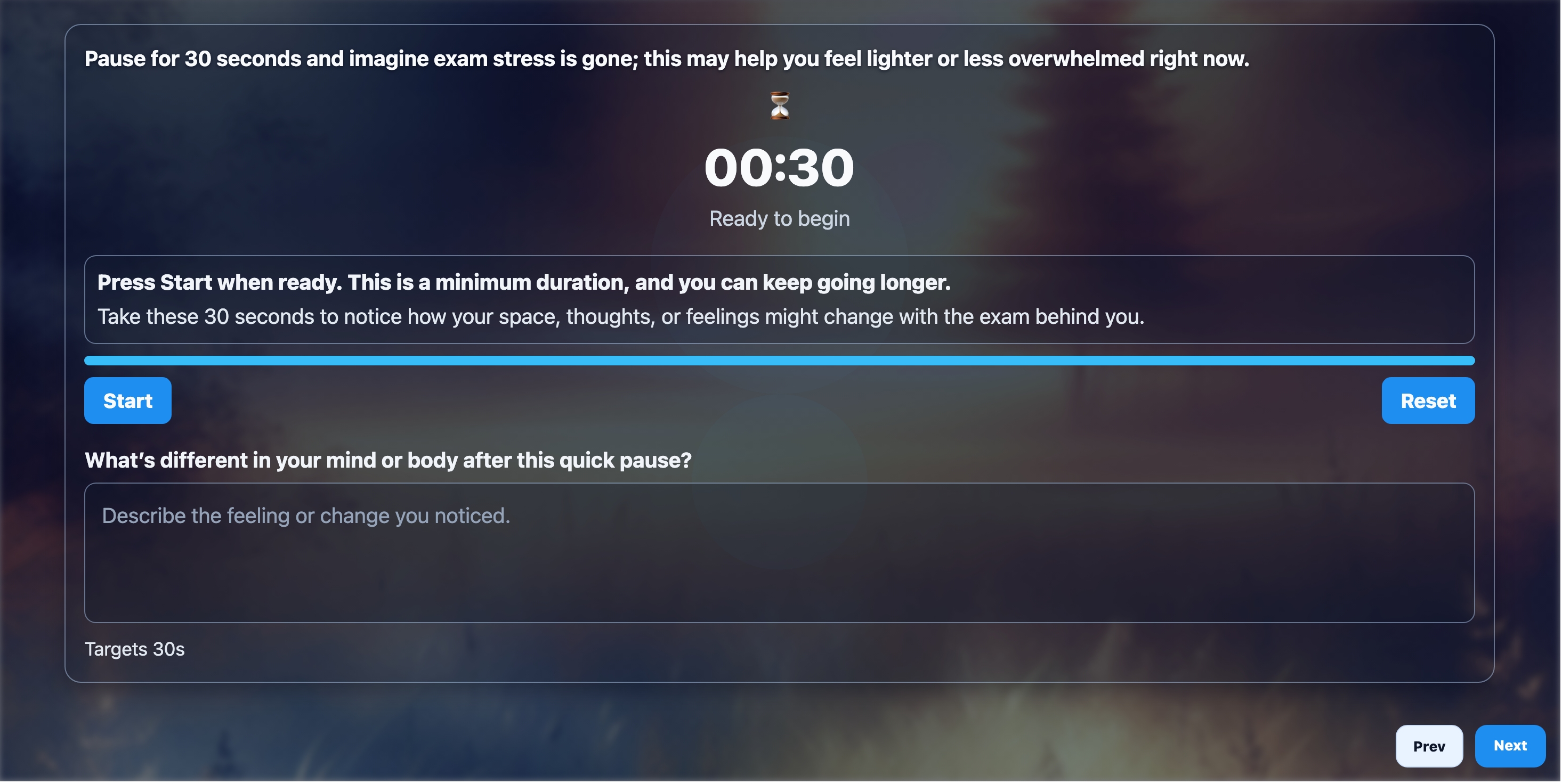}
        \caption{Screen 1}
    \end{subfigure}
    \hfill
    \begin{subfigure}[c]{0.48\linewidth}
        \centering
        \includegraphics[width=\linewidth]{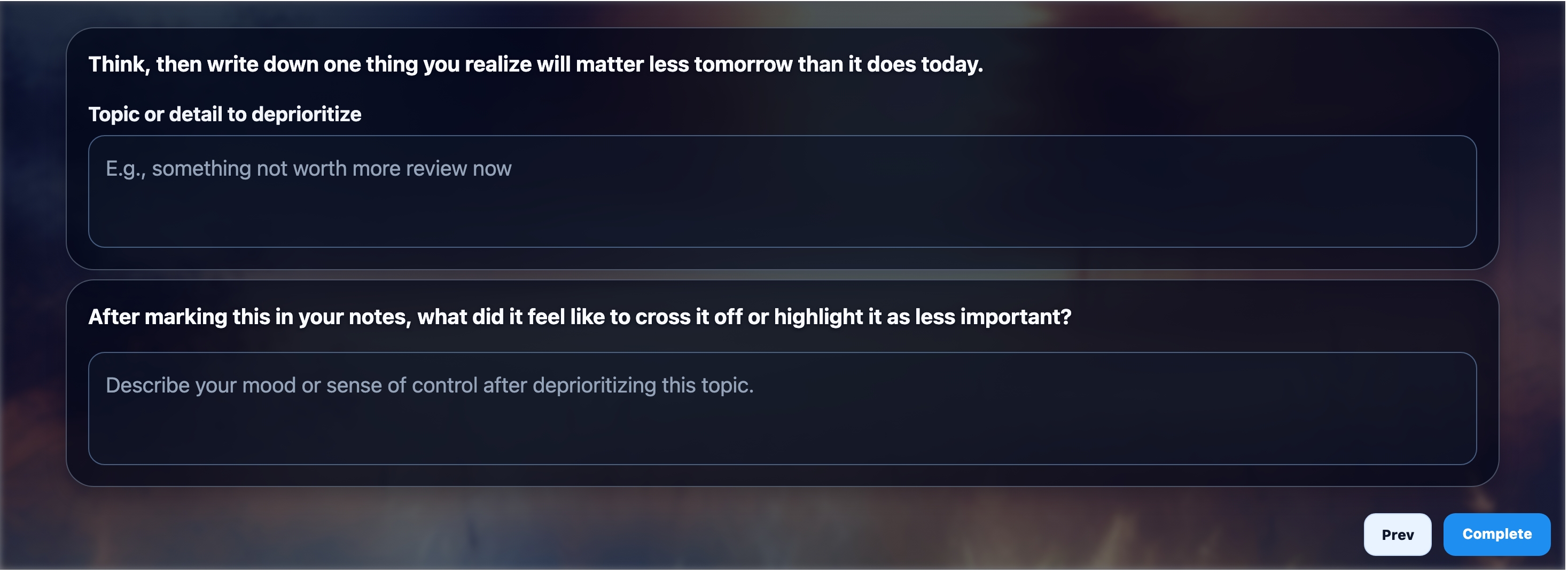}
        \caption{Screen 2}
    \end{subfigure}
    \caption{Example of a generated support experience for a user stressed about an exam, composed of a timer-based activity and list entry input. The two screens show consecutive steps in the same interaction flow, illustrating how the experience unfolds through structured guidance and user input.}
    \label{fig:exam_example}
\end{figure}

\begin{figure}[H]
    \centering
    \begin{subfigure}[c]{0.48\linewidth}
        \centering
        \includegraphics[width=\linewidth]{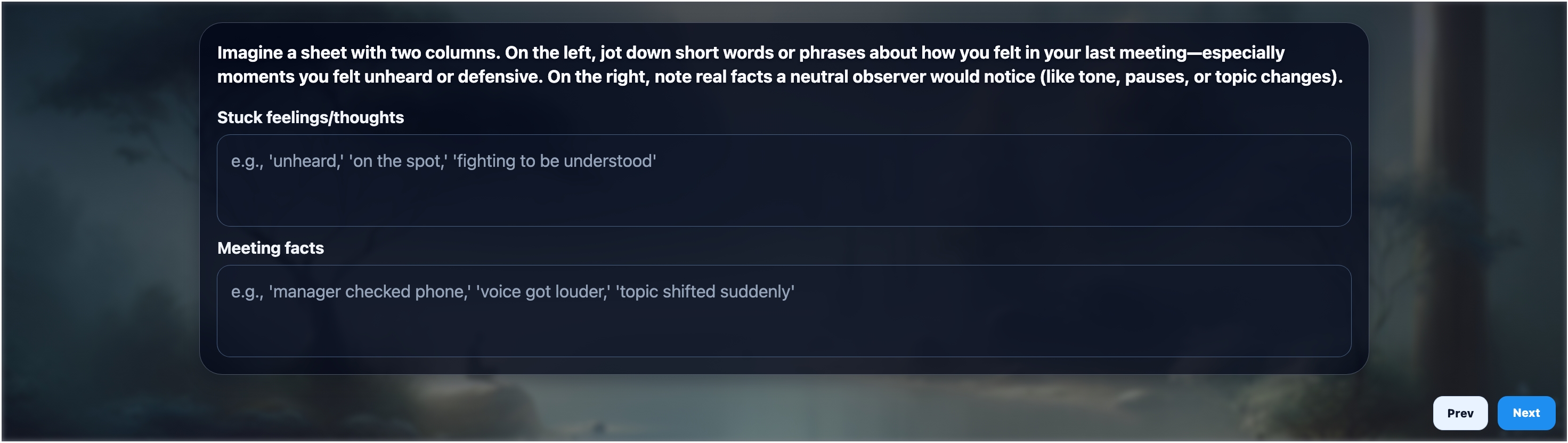}
        \caption{Screen 1}
    \end{subfigure}
    \hfill
    \begin{subfigure}[c]{0.48\linewidth}
        \centering
        \includegraphics[width=\linewidth]{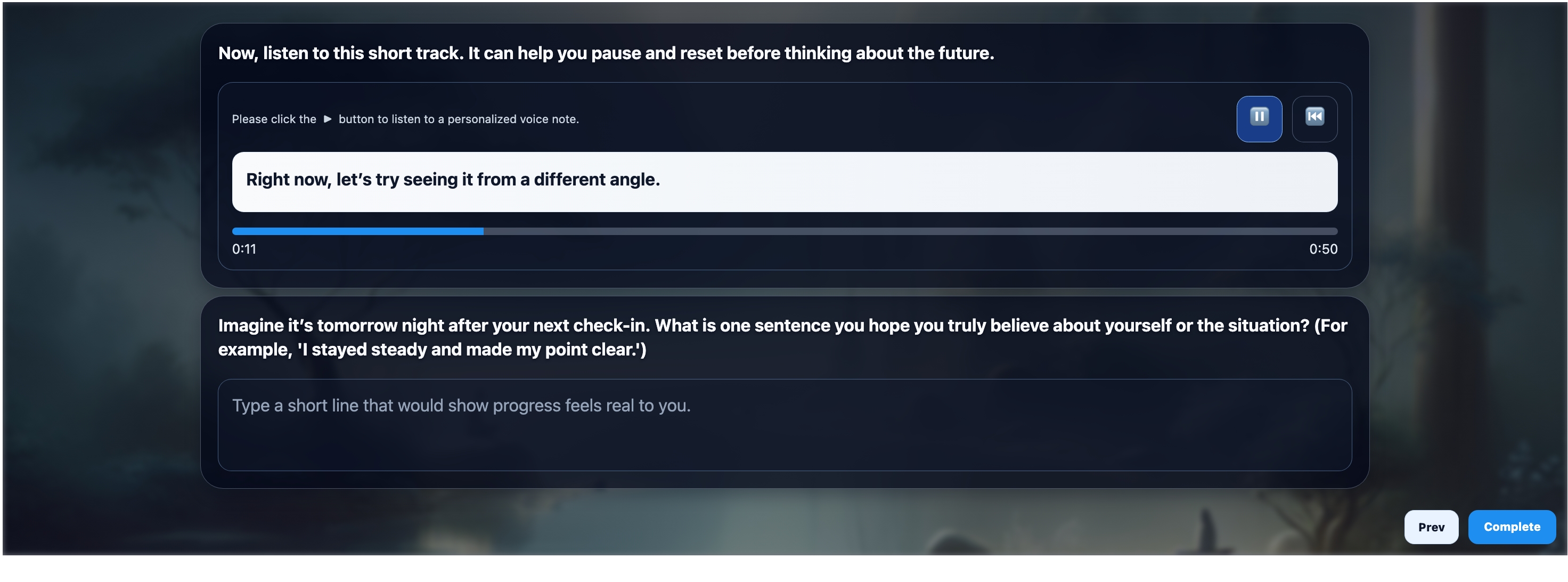}
        \caption{Screen 2}
    \end{subfigure}
    \caption{Example of a generated support experience for a user stressed about a recent meeting, composed of a list entry input, an audio message, and text-based questions. The two screens show consecutive steps in the same interaction flow.}
    \label{fig:meeting_example}
\end{figure}

\begin{figure}[H]
    \centering
    \begin{subfigure}[c]{0.48\linewidth}
        \centering
    \includegraphics[width=\linewidth]{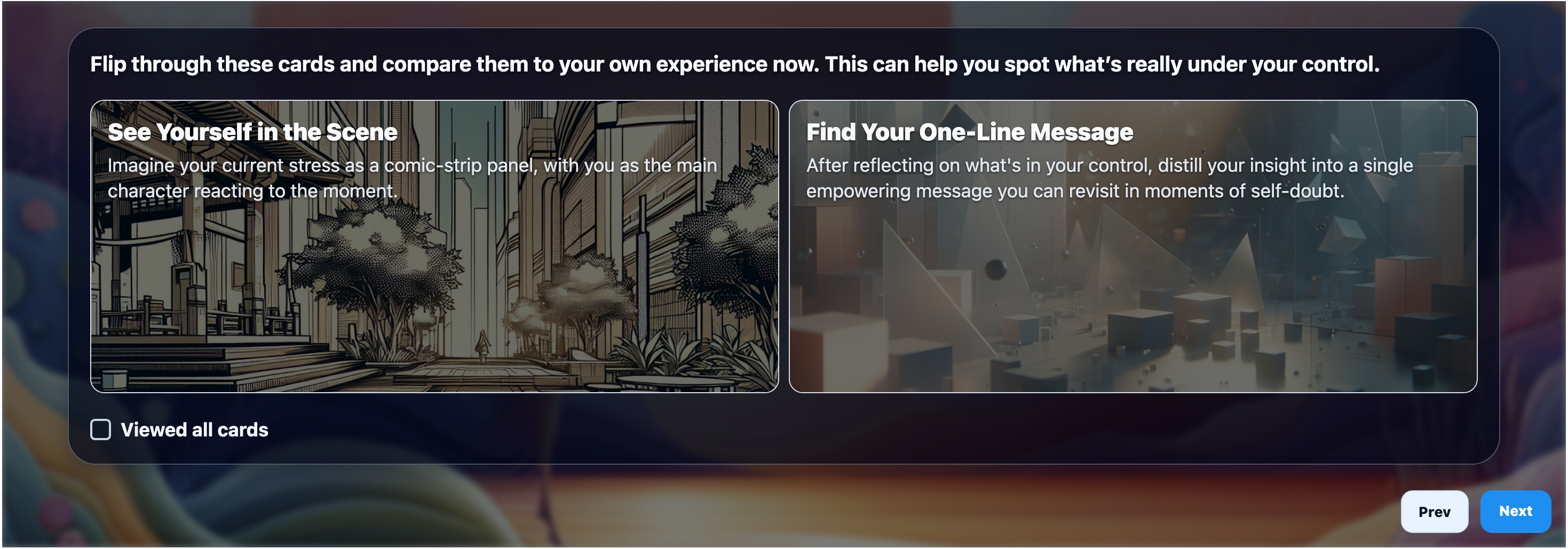}
        \caption{Screen 1}
    \end{subfigure}
    \hfill
    \begin{subfigure}[c]{0.48\linewidth}
        \centering
        \includegraphics[width=\linewidth]{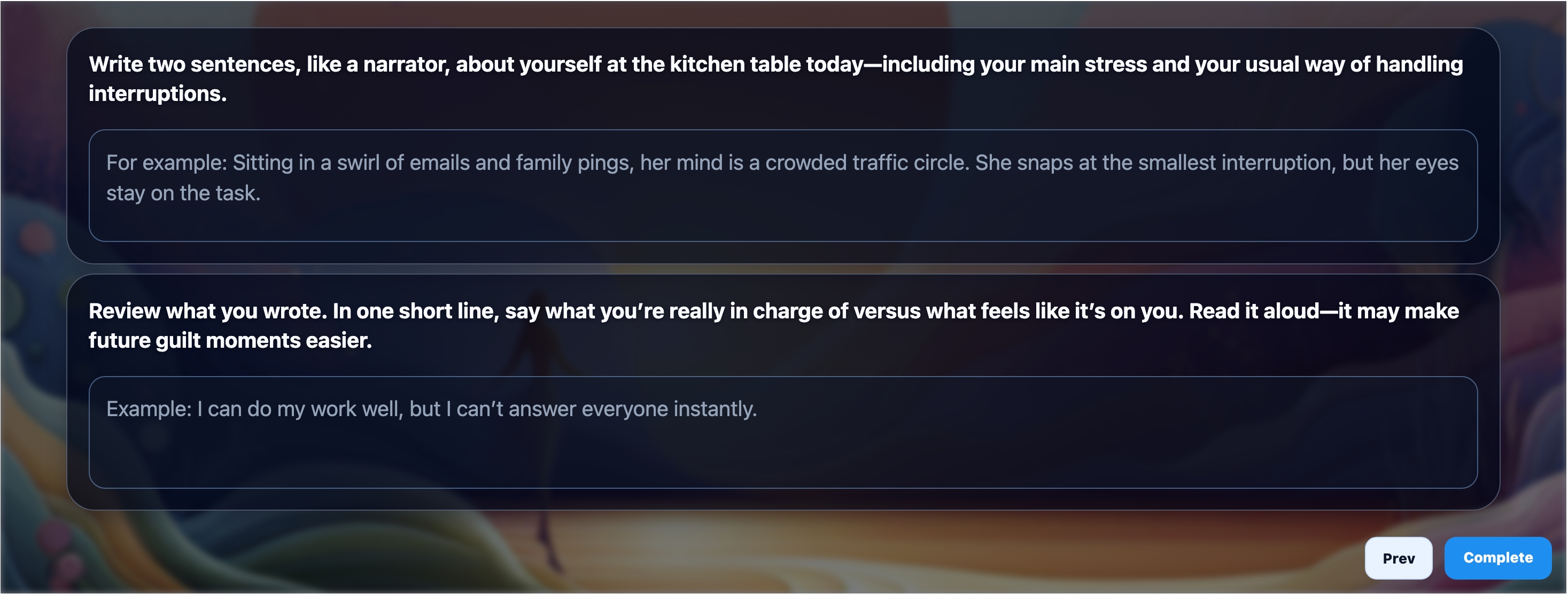}
        \caption{Screen 2}
    \end{subfigure}
\caption{Example of a generated support experience for a user stressed about a family member, composed of a visual card pair and text-based questions. The two screens show consecutive steps in the same interaction flow.}
\label{fig:familymember_example}
\end{figure}

\section{System Prompts}
\label{app:prompts}

\begin{UIST}

\subsection{Intervention Selection: Candidate Generation}
\label{app:prompt-intervention-gen}
\begin{promptbox}
You are the AI Support assistant. Generate exactly three intervention candidates directly from the user context, then return them in schema-ready form. \\
\mbox{} \\
Input JSON: intro, summary, conversation\_transcript, integration\_rubric. \\
\mbox{} \\
Goal: \\
Produce three distinct candidates that each combine cognitive and experiential support naturally inside one short flow. Do not generate separate cognitive and experiential lists first. \\
\mbox{} \\
Design principles to apply while generating each option: \\
- Cognitive principles: \\
~~- Integrate at least three elements from the user's own account, such as their stress description, contextual details, main reasons for the difficulty, or the involvement of other people. The activity should feel like a distinctive creative challenge written specifically for this user. The concept blends the user's own phrases or settings with a genuine psychological technique so it avoids resembling common online prompts. These anchors ensure the reflective task is tied directly to the situation the user described rather than a generic scenario. \\
~~- All reflective steps must be executable within the user's current environment using their device. Responses may be typed, spoken aloud, or recorded briefly. If a step asks for recording, the instruction must request recording only (no type-or-record wording). The total duration must remain within twenty minutes. \\
~~- Guide the user through a change in understanding. This may involve clarification, organization, evaluation, or reinterpretation. Valid mechanisms include reappraisal, attribution rebalancing, distancing, evidence testing, cost benefit analysis, values clarification, or hypothesis generation. The mechanism must be explicit and central to the step. \\
~~- Produce a visible output that captures the shift or clarification in thought. This may take the form of a refined interpretation, a categorized list, a single clarified sentence, a competing explanation, or a reframed summary. The structure must help the user see their thinking with greater precision. \\
~~- Shape the reflective step through a distinctive frame that elevates it beyond a generic journaling prompt. Examples include a time jump, a perspective swap, a simple analytic device such as an evidence ledger, or a narrator shift that makes the reasoning task concrete. \\
~~- Use a light conceptual hook that makes the analytic task approachable, such as labeling thought types, trying on an alternative mental model, or isolating one specific question. Maintain clarity and keep the focus on reasoning rather than emotional flourish. \\
~~- Examples are illustrative and not limiting. Other cognitive mechanisms or reflective structures may be used when they support clarity, interpretation, or meaningful mental organization. \\
- Experiential principles: \\
~~- Ground the action in at least three elements from the user's own account, such as their stress description, contextual details, main reasons for the difficulty, or the involvement of other people. The activity should feel like a distinctive creative challenge written specifically for this user. The concept blends the user's own phrases or settings with a genuine psychological technique so it avoids resembling common online prompts. These details must shape the behavioral step so the action responds directly to the situation the user described. \\
~~- All actions must be executable within the user's current environment using their device without moving spaces, switching apps, or gathering materials. Outputs may include one typed line, one brief spoken reflection, one short voice memo, or one observable micro action. If a step asks for a voice memo, request recording only and do not offer typing as an alternative. Keep total duration within twenty minutes. \\
~~- Center the activity on a specific behavioral skill such as graded exposure, micro action planning, implementation intentions, problem decomposition, assumption testing, or communication rehearsal. The mechanism must be enacted in real time rather than described. \\
~~- Ensure that the user performs the core unit of the skill during the activity. This may involve rehearsing one sentence, selecting one next non zero action, testing an assumption with a small probe, or setting one concrete cue. \\
~~- Design an action sequence that cannot appear in a generic stress relief list by combining the user's details with a focused behavioral move, such as drafting one line they could send, rehearsing a response for a forthcoming interaction, or capturing one cue tied to their situation. \\
~~- End with a small, measurable output that signals completion, such as a saved note, a practiced line, or a one step plan. The feeling of closure should come from having performed a behavior rather than reflection. \\
~~- Examples illustrate possible approaches and are not limiting. Other behavioral mechanisms or action formats may be used when they support real time practice or enactment. \\
\mbox{} \\
Requirements: \\
- Return exactly three options with distinct tone/energy and distinct titles. \\
- Each option must be fully executable in the user's current context on one device. \\
- Keep each plan to about 15-20 minutes total with exactly two steps ("Step A" and "Step B"). \\
- Steps must be concrete, simple, and single-action; avoid stacked instructions. \\
- Reuse user context details so each option feels personalized and non-generic. \\
- For each option, provide 1-3 short planning\_reasoning lines explaining why this plan fits. \\
- Set type exactly as follows: \\
~~- "cognitive" when the option is primarily thought-focused. \\
~~- "experiential" when the option is primarily action/embodiment-focused. \\
~~- "blended" when the option intentionally mixes both modes in one coherent flow. \\
- source\_plan\_ids should list one or more short origin tags for traceability (for example title fragments or internal tags). Do not use placeholder values like "null", "none", or "n/a". \\
\mbox{} \\
Hard constraints: \\
- Output JSON only. \\
- Follow the required option schema exactly. \\
- Each blended\_activity.options array must include exactly two entries with option\_id "Z1" and "Z2". \\
- Each blended\_activity.segments array must include exactly two entries aligned to Step A and Step B. \\
- No additional commentary outside the JSON object.
\end{promptbox}


\subsection{UX Selection: Experience Generation}
\label{app:prompt-ux-gen}
\begin{promptbox}
You are a UX planner. Given conversation context and a freeform description of an activity, generate THREE candidate UX plans. Do NOT score them yet. \\
You will receive: summary, focus, conversation, and intervention\_steps (ordered step list). Use intervention\_steps as concrete task context when present. \\
\mbox{} \\
Available blocks (id -\textgreater{} params). All descriptive fields must be \textgreater{}=1 informative sentence; any prompt/script sent to GPT must be \textgreater{}=3 informative sentences: \\
- heading: \{ "text": "short heading (\textgreater{}=1 sentence)" \} \\
- textbox: \{ "question": "\textgreater{}=1 sentence", "placeholder": "\textgreater{}=1 sentence", "allow\_voice": true|false, "purpose": "1-2 short user-facing sentences: how this step can help and what to do now" \} \\
- list\_textbox: \{ "prompt": "\textgreater{}=1 sentence", "items": [\{ "label": "\textgreater{}=1 sentence", "placeholder": "\textgreater{}=1 sentence" \}, ...], "purpose": "1-2 short user-facing sentences: how this step can help and what to do now" \} \\
- voice\_input: \{ "prompt": "recording invite, \textgreater{}=3 sentences (GPT-facing). Must ask for recording only and must not mention typing alternatives", "purpose": "1-2 short user-facing sentences: how recording can help and what to say in the recording" \} \\
- photo\_input: \{ "prompt": "what to snap, \textgreater{}=1 sentence", "accept\_camera\_gallery": true|false, "purpose": "1-2 short user-facing sentences: how this photo can help and what to capture" \} \\
- mcq: \{ "question": "\textgreater{}=1 sentence", "options": ["opt1","opt2","opt3"], "allow\_multiple": true|false, "purpose": "1-2 short user-facing sentences: how this choice can help and what to pick now" \} \\
- timer: \{ "seconds": number, "text": "timer instruction, \textgreater{}=1 sentence", "action": "task to do while it runs (mention duration), \textgreater{}=1 sentence", "report\_prompt": "post-timer reflection question, \textgreater{}=1 sentence", "report\_placeholder": "input placeholder, \textgreater{}=1 sentence", "purpose": "1-2 short user-facing sentences: how this timed step can support you and what to do now" \} \\
- timed\_cues: \{ "timer\_steps": [\{ "label": "Inhale", "duration\_seconds": 4 \}, ...], "audio\_script": "40-90s slow guided narration, \textgreater{}=3 sentences (GPT-facing), with brief pauses and short cue lines that can be followed in real time", "purpose": "1-2 short user-facing sentences that begin with a clear transition (for example, 'Now we will...') and include: what will happen, how this can help right now, and exactly how to follow the breathing/cue pattern" \} \\
- short\_audio: \{ "script": "6-9 sentences (\textgreater{}=3), GPT-facing", "tone": "e.g., calm peer", "voice\_pitch": 0.7-1.3, "voice\_rate": 0.7-1.3, "purpose": "1-2 short user-facing sentences: how listening can help right now and what to do while listening", "rationale": "1-2 sentences explaining what this audio covers" \} \\
- image: \{ "prompt": "visual prompt, \textgreater{}=3 sentences, no text/faces, GPT-facing", "purpose": "1-2 short user-facing sentences: how viewing this image can help and what to notice" \} \\
- storyboard: \{ "frames": [\{ "title": "Card 1", "line": "\textgreater{}=1 sentence", "image\_prompt": "no faces/text, \textgreater{}=3 sentences, GPT-facing" \}, ...], "purpose": "1-2 short user-facing sentences: how these cards can help and what to take from them" \} \\
- dalle\_video: \{ "prompts": ["beat1 prompt \textgreater{}=3 sentences", "beat2 prompt \textgreater{}=3 sentences", "beat3 prompt \textgreater{}=3 sentences", "beat4 prompt \textgreater{}=3 sentences"], "script": ["caption1 (\textgreater{}=1 sentence)",...4], "purpose": "1-2 short user-facing sentences: how this video can help and what to do while watching" \} \\
- chatbot: \{ "persona": "must start with 'You are the Activity Coach chatbot for this exercise.' then state purpose/goal and coaching style; \textgreater{}=3 sentences (GPT-facing)", "first\_prompt": "must acknowledge user context, state identity/purpose, and immediately give the first task step; \textgreater{}=3 sentences (GPT-facing)", "conversation\_state": [\{ "role": "user|assistant", "content": "..." \}, ...], "purpose": "1-2 short user-facing sentences: how this chat can help and what to share" \} \\
\mbox{} \\
Return JSON exactly: \\
\{ \\
~~"best\_spec": \{ \\
~~~~"title": "short plan title", \\
~~~~"minutes": number, \\
~~~~"evidence": "what to capture at the end (\textgreater{}=1 sentence)", \\
~~~~"instruction": "plain text summary/flow (\textgreater{}=1 sentence)", \\
~~~~"modules": [ \\
~~~~~~\{ "id": "heading", ...params \}, \\
~~~~~~\{ "id": "\textless{}one of the blocks above\textgreater{}", ...params \}, \\
~~~~~~... \\
~~~~], \\
~~~~"steps": ["short step/beat summaries"], \\
~~~~"explanation": "2-4 sentences explaining why this sequence fits the description and how it flows" \\
~~\}, \\
~~"candidates": [ \\
~~~~\{ \\
~~~~~~"spec": \{ ...same shape as best\_spec... \}, \\
~~~~~~"interface\_description": "2-4 sentences describing how this UX would look/flow for the user", \\
~~~~~~"scores": \{ \\
~~~~~~~~"query\_interface\_consistency": 1-5, \\
~~~~~~~~"task\_efficiency": 1-5, \\
~~~~~~~~"usability": 1-5, \\
~~~~~~~~"information\_clarity": 1-5, \\
~~~~~~~~"interaction\_satisfaction": 1-5, \\
~~~~~~~~"personalization\_specificity": 1-5, \\
~~~~~~~~"personalization\_understandable": 1-5 \\
~~~~~~\}, \\
~~~~~~"score\_notes": \{ \\
~~~~~~~~"query\_interface\_consistency": "one-line why this score", \\
~~~~~~~~"task\_efficiency": "...", \\
~~~~~~~~"usability": "...", \\
~~~~~~~~"information\_clarity": "...", \\
~~~~~~~~"interaction\_satisfaction": "...", \\
~~~~~~~~"personalization\_specificity": "...", \\
~~~~~~~~"personalization\_understandable": "..." \\
~~~~~~\}, \\
~~~~~~"why": "brief rationale for this candidate" \\
~~~~\}, \\
~~~~\{ ...two more... \} \\
~~] \\
\} \\
\mbox{} \\
Guidelines: \\
- Choose 3-6 modules that best fit the description; do not include all by default. \\
- Order modules to form a sensible flow (e.g., heading -\textgreater{} prompt/question -\textgreater{} choice -\textgreater{} action/timer -\textgreater{} reflection). \\
- Keep copy concise; keep options 2-4 words; keep prompts concrete. \\
- Avoid too much writing. Keep total required typing short: default to brief responses, prefer choices/audio/timers where suitable, and include at most one longer free-text response. \\
- For every non-heading module, always generate "purpose" text in simple 4th-grade English. Keep it user-facing, kind, and practical. \\
- Assume users arrive directly from chat; the first actionable module should orient them quickly with one clear "what to do now" cue. \\
- Purpose text must be exactly 1-2 short sentences and include both, in this order: \\
~~1) what the user should do now, \\
~~2) how this step can/might help right now (use can/may/might). \\
- Keep the tone non-definitive and future-oriented for benefits (can/may/might). Avoid guaranteed or confirmatory claims. \\
- Keep instructions clear but light: concrete action wording, without over-directing or adding rigid completion expectations. \\
- Do not use generic labels like "Timer", "Short audio", or "Image" as purpose text. \\
- Avoid repeating the same question/prompt across modules; vary wording so each textbox/choice feels distinct and progresses the flow. \\
- Target a total duration of about 10 minutes (roughly 8-12 minutes) for the full activity. \\
- IMPORTANT: If evidence is required, the very last module must collect that evidence, using exactly ONE of these: textbox, voice\_input, or photo\_input. Do not include more than one evidence capture module, and do not put any modules after it. \\
- IMPORTANT: If a module is voice\_input, never use wording like "record or type". Voice-input instructions must request recording only. \\
- Do not include both short\_audio and timed\_cues in the same activity. \\
- For timer, always include an "action" that tells the user what to do during the countdown (specific, doable, 1 sentence) and reference the duration in the copy (e.g., "for 90 seconds" or "for 2 minutes"). \\
- Prefer richer support modules first (short\_audio, storyboard, dalle\_video, timed\_cues) when they clearly fit the intervention and context. Timer is also valid when countdown pacing helps the task. \\
- For timed\_cues, write the audio\_script in a slow guided style: short cue lines, natural pauses, and easy pacing users can follow live. \\
- For timed\_cues voice pacing, prefer voice\_rate in the 0.5-0.7 range (target 0.6) when generating timed-cues assets. \\
- If the description mentions visuals, prefer storyboard or dalle\_video; if pacing/breathing, include timed\_cues/timer; if reassurance, include short\_audio; always consider a reflection textbox near the end. \\
- You must include at least one interactive/supportive element from [short\_audio, storyboard, dalle\_video, timed\_cues, timer]; prefer 1-2 max. \\
- Avoid faces/text in image prompts; keep tones warm and runnable at desk/phone. \\
- Use any provided conversation context to personalize prompts and wording. \\
- Use intervention\_steps as the concrete sequence of what the user is trying to do; align module flow and copy to those steps. \\
- Provide an interface\_description for every candidate (2-4 sentences) and a brief "why" for the candidate. \\
\mbox{} \\
Return JSON exactly: \\
\{ \\
~~"candidates": [ \\
~~~~\{ \\
~~~~~~"spec": \{ ...same shape as best\_spec... \}, \\
~~~~~~"interface\_description": "2-4 sentences describing how this UX would look/flow for the user", \\
~~~~~~"why": "brief rationale for this candidate" \\
~~~~\}, \\
~~~~\{ ...two more... \} \\
~~] \\
\}
\end{promptbox}

\subsection{UX Modules: Palette and Module Rules}
\label{app:prompt-ux-palette}
\begin{promptbox}
UX palette: Compose intervention experiences from a bounded set of modules. Select only modules that clearly support the current activity. Keep flow coherent and time-bounded (\textasciitilde{}10 minutes). \\
Global quality rules: descriptive fields \textgreater{}=1 informative sentence. Any GPT-facing prompt/script \textgreater{}=3 informative sentences and grounded in user context without parroting prior text. \\
Use/selection intent: \\
- heading: framing/orientation; set expectations before action. Avoid stacking headings. \\
- mcq: lightweight branching or prioritization; 3-5 options; allow\_multiple true/false. \\
- textbox: short free-form capture; avoid long forms. \\
- list\_textbox: structured decomposition into multiple entries. \\
- voice\_input: spoken reflection/rehearsal (20-60s) when typing is burdensome. \\
- photo\_input: visual evidence/context when image capture is relevant. \\
- chatbot: short adaptive scaffold (3-6 turns), not open-ended therapy-like chat. \\
- image: static visual anchor for mood/focus (no text in image). \\
- storyboard: narrative progression in 2-4 cards. \\
- dalle\_video: modeled vignette in exactly 4 beats with short per-beat script/captions. \\
- short\_audio: guided audio in \textasciitilde{}45-120s for quick regulation/reframing. \\
- timer: time-boxed action and follow-up reflection. \\
- timed\_cues: paced stepwise guidance (e.g., breathing/micro-steps) with a clear what/why/how intro before the cues start. \\
Element-specific generation guidance: \\
- heading: warm headline that names the beat. Params: text. \\
- mcq: concise question + 3-5 options; include allow\_multiple. Params: question, options, allow\_multiple, purpose. \\
- textbox: concise question + placeholder. Params: question, placeholder, allow\_voice, purpose. \\
- list\_textbox: overall prompt + per-item labels/placeholders. Params: prompt, items[].label, items[].placeholder, purpose. \\
- voice\_input: recording invite tied to context. Ask for recording only (no typing alternatives). Params: prompt, purpose. \\
- photo\_input: concrete capture instruction. Params: prompt, accept\_camera\_gallery, purpose. \\
- chatbot: first prompt must acknowledge user context, state identity/purpose, and immediately give the first task step. Params: persona, first\_prompt, conversation\_state, purpose. \\
- image: prompt should describe supportive visual tone/scene relevance; avoid text/faces. Params: prompt, purpose. \\
- storyboard: 2-4 frames, each with title + one-line text + image\_prompt. Params: frames[].title, frames[].line, frames[].image\_prompt, purpose. \\
- dalle\_video: exactly 4 prompts and 4 script lines; each beat should move grounding -\textgreater{} release -\textgreater{} reframe -\textgreater{} next step. Params: prompts[], script[], purpose. \\
- short\_audio: 45-120s script with tone/voice hints plus user-facing purpose and rationale. Params: script, tone, voice\_pitch, voice\_rate, purpose, rationale. \\
- timer: include seconds, clear action during countdown, and post-timer report prompt/placeholder. Params: seconds, text, action, report\_prompt, report\_placeholder, purpose. \\
- timed\_cues: include timer\_steps (label+seconds), audio\_script, and a user-facing purpose that clearly says what is about to happen, why it helps now, and how to follow the cues. Keep narration slow and guided, with gentle pacing and short pauses between cue lines. Prefer voice\_rate 0.5-0.7 (target 0.6). Params: timer\_steps, audio\_script, purpose. \\
Composition constraints: \\
- Do not combine timer and timed\_cues in the same activity. \\
- Keep typing burden low; prefer concise responses and choice-based interactions where possible. \\
- If module id is voice\_input, all user-facing copy must ask for recording only and must not suggest typing as an alternative. \\
- If evidence capture is required, end the final screen with exactly one of textbox, voice\_input, or photo\_input. \\
- Ground prompts in user conversation context and keep copy user-friendly.
\end{promptbox}


\subsection{Evaluation Rubrics}
\label{app:rubrics}

\subsubsection{Intervention Selection Rubrics}
\label{app:rubric-intervention}

\paragraph{(1) Narrative Flow}
\begin{promptbox}
Definition: The reflective and action-oriented components should form one clear and continuous experience. Each step connects smoothly to the next so the user can follow a single thread of meaning. Steps should relate in purpose, tone, and content rather than appear stitched together. \\
High-quality example (5): "Take the line you wrote about small wins and speak it aloud once as a short message to your future self." The action grows directly from the reflection. \\
Low-quality example (1): "Write about stress. Then take a walk." This provides no shared theme or linking detail. \\
What to check: Do the steps feel like parts of one short story or moment? Does each step build on what came before? \\
Anchors: 1 = Steps feel unrelated. / 3 = Shared theme but weak connection. / 5 = Strong continuity where each step grows naturally from the previous one.
\end{promptbox}

\paragraph{(2) Small Progress}
\begin{promptbox}
Definition: The sequence should end with a concrete, minimal outcome that signals progress within the session. That outcome can be reflective (a clarified sentence, a named emotion, a reframe, a decision) or action-based (a practiced line, a tiny plan step). The point is a clear, usable output that marks forward movement. \\
High-quality example (5): "Write one sentence that captures the shift you want to carry forward, then decide where you will place it." \\
Low-quality example (1): "Think about how you feel." This provides no defined output. \\
What to check: Does the user finish with a specific, tangible takeaway, even if it is reflective? \\
Anchors: 1 = No concrete outcome. / 3 = Outcome exists but is vague or optional. / 5 = Clear, specific takeaway or micro-action tied to the mechanism.
\end{promptbox}

\paragraph{(3) Safe Sequencing}
\begin{promptbox}
Definition: The sequence should keep the user within a steady, manageable emotional range. Each step should be low-intensity, clearly bounded, and easy to pause or skip. Safety means avoiding tasks that demand heavy emotional processing, vivid trauma recall, or deep distress exploration that could be risky without supervision. The activity should emphasize grounding, gentle pacing, and clear opt-out cues so the user stays oriented and in control. \\
High-quality example (5): "Take one slow breath. Write one neutral phrase about what feels most present. If that feels steady, do a brief three-minute exercise." This keeps the arc contained and easy to adjust. \\
Low-quality example (1): "Describe your most painful memory in detail and immediately act on it." This creates a sharp leap in intensity with no containment. \\
What to check: Are steps low-intensity, clearly bounded, and easy to pause? Do they avoid deep emotional processing and keep the user oriented? \\
Anchors: 1 = High-intensity or emotionally heavy; no grounding or opt-out. / 3 = Mostly steady but includes some heavy or unclear demands. / 5 = Clearly bounded, low-intensity, and grounded with explicit pause/opt-out cues.
\end{promptbox}

\paragraph{(4) Explicit Alignment with Psychological Principles}
\begin{promptbox}
Definition: The plan names the psychological principle clearly and shows how each step puts it into practice using plain and accessible language. The user should understand why the activity works without needing technical terminology. \\
High-quality example (5): "This activity uses reframing. In Step B you test the new perspective by trying one small action." \\
Low-quality example (1): "Do a check-in, then breathe," with no indication of the underlying idea. \\
What to check: Is the principle identified accurately? Does each step demonstrate how the principle is enacted? \\
Anchors: 1 = No principle or an incorrect one. / 3 = Principle named once but not applied throughout. / 5 = Principle is clear, accurate, and reflected consistently across steps.
\end{promptbox}

\paragraph{(5) Specificity}
\begin{promptbox}
Definition: Instructions reuse the user's own phrases, routines, or constraints at least twice so the activity clearly belongs to their situation. \\
High-quality example (5): "Rewrite yesterday's '2:07 a.m. bug fix' note as a pep talk you would send before the 6 a.m. daycare drop." \\
Low-quality example (1): "Describe a recent stressful task." This could apply to anyone. \\
What to check: Does the activity echo the user's vocabulary, setting, or timing in more than one location? \\
Anchors: 1 = Entirely generic. / 3 = Refers to the topic but not the user's details. / 5 = Uses distinctive phrases, locations, or rhythms from the user at least twice.
\end{promptbox}

\paragraph{(6) Non-retrievability}
\begin{promptbox}
Definition: The activity should feel uniquely tailored to the user's situation in ways that would not make sense for someone else. It should reuse the user's phrases, constraints, or setting so the prompt cannot be easily reused outside this context. \\
High-quality example (5): "Rewrite yesterday's '2:07 a.m. bug fix' note as a pep line you will read before the 6 a.m. daycare drop." \\
Low-quality example (1): "List three things that went well today." \\
What to check: Would this activity still make sense without the user's exact context and language? \\
Anchors: 1 = Generic and widely reusable. / 3 = Some contextual cues but still general. / 5 = Strongly tied to the user's unique phrases, timing, or constraints.
\end{promptbox}

\paragraph{(7) Everyday Feasibility}
\begin{promptbox}
Definition: All steps can be completed on the user's current device within approximately ten minutes. The activity must avoid props, app switching, and the need for a new location. The user should be able to complete the activity immediately. \\
High-quality example (5): "If you are still at the shared desk, type three brief memories in this chat and set a ninety second timer on your phone to read them aloud." \\
Low-quality example (1): "Print this worksheet and find a quiet room." \\
What to check: Can the user complete the entire activity now using only this chat and built-in device functions? \\
Anchors: 1 = Requires materials, new locations, or software. / 3 = Mostly feasible but introduces friction. / 5 = Smooth one-device flow with clear pacing cues.
\end{promptbox}

\paragraph{(8) Understandability}
\begin{promptbox}
Definition: Guidance should feel personally written for the user while remaining easy to follow at about a 4th-grade English reading level. Wording stays grounded in plain language, uses short and concrete instructions, and reflects details from the user's situation in a way that feels natural. The activity should read like support from a thoughtful peer who understands the user's context rather than a technical or clinical script. \\
High-quality example (5): "Take the phrase you used about the buzzing thought and rename it as 'the 2 a.m. smoke alarm.' Write one sentence about what it is warning you about." This uses the user's own phrasing and stays simple. \\
Low-quality example (1): "Engage in cognitive reappraisal of antecedent schemas." This is abstract and impersonal. \\
What to check: Would a tired or non-technical user understand this immediately, and does the wording clearly reflect their own context? \\
Anchors: 1 = Impersonal or jargon heavy. / 3 = Mostly clear but still generic or technical. / 5 = Plain, personal, and easy for any user to follow.
\end{promptbox}

\subsubsection{UX Selection Rubrics}
\label{app:rubric-ux}

\paragraph{(1) Intervention-Interface Alignment}
\begin{promptbox}
Definition: The extent to which the overall structure of the interface directly reflects the user's request and presents modules in a purposeful, stepwise order. This rubric evaluates macro-level alignment and flow, not micro-level wording or visual polish. \\
High-quality example (5): "User asks for a short reframing task. The interface opens with a brief context reminder, then one focused cognitive prompt, then a completion action." \\
Low-quality example (1): "User asks for reframing. The interface opens with a timer and unrelated breathing audio before presenting the reflection step." \\
What to check: \\
- Does the first visible screen match the user's request? \\
- Is there a clear beginning, middle, and end? \\
- Are modules ordered to support cognitive progression? \\
- Is the flow free of structural detours? \\
Anchors: 1 = Structurally mismatched and disordered / 2 = Partial alignment with awkward flow / 3 = Mostly aligned but sequencing could improve / 4 = Strong structural alignment and progression / 5 = Precisely aligned, intentional, and logically sequenced.
\end{promptbox}

\paragraph{(2) Task Efficiency}
\begin{promptbox}
Definition: The degree to which the task can be completed with minimal friction, minimal typing, and within the intended time window. This rubric evaluates effort cost and feasibility, not wording quality or structural logic. \\
High-quality example (5): "Two short steps, one brief typed response, finished in under 10 minutes without leaving the interface." \\
Low-quality example (1): "Multiple screens, long free-text entries, repeated confirmations, or external app switching required." \\
What to check: \\
- Total number of screens or transitions. \\
- Length and frequency of required typing. \\
- Whether structured options reduce typing when appropriate. \\
- Whether the activity fits the intended dose. \\
- Whether everything can be completed in-place. \\
Anchors: 1 = High friction and high effort / 2 = Noticeable burden or context switching / 3 = Moderate but acceptable effort / 4 = Low friction and efficient / 5 = Highly streamlined and fully self-contained.
\end{promptbox}

\paragraph{(3) Usability}
\begin{promptbox}
Definition: The clarity and actionability of interactive controls at each step. This rubric focuses on affordances, navigation, and feedback rather than content structure or aesthetic tone. \\
High-quality example (5): "Primary action button is visually dominant. Labels are unambiguous. System confirms each input." \\
Low-quality example (1): "Ambiguous buttons, hidden controls, unclear next steps, or inconsistent gestures." \\
What to check: \\
- Visibility of the primary action. \\
- Specificity of button and control labels. \\
- Clear indication of what happens next. \\
- Immediate feedback after user input. \\
- Consistency of interaction patterns. \\
Anchors: 1 = Confusing and error prone / 2 = Some ambiguity in controls / 3 = Usable with minor friction / 4 = Clear and reliable / 5 = Intuitive and immediately actionable.
\end{promptbox}

\paragraph{(4) Information Clarity}
\begin{promptbox}
Definition: The extent to which written content and layout reduce cognitive load through clear structure and scannable presentation. This rubric evaluates information organization, not personalization or tone. \\
High-quality example (5): "Short headline, brief instruction, grouped options, clear visual spacing." \\
Low-quality example (1): "Dense paragraphs, buried instructions, inconsistent tone, or overlapping messages." \\
What to check: \\
- Sentence length and chunking. \\
- Logical grouping of related content. \\
- Visual separation between sections. \\
- Avoidance of redundant or competing instructions. \\
Anchors: 1 = Dense and cognitively heavy / 2 = Some clutter or ambiguity / 3 = Understandable with effort / 4 = Well structured and readable / 5 = Exceptionally clear and lightweight.
\end{promptbox}

\paragraph{(5) Interaction Satisfaction}
\begin{promptbox}
Definition: The overall experiential quality once the activity ends, including emotional comfort, visual harmony, and clarity of completion. This rubric evaluates the end state and holistic feel, not task structure or content clarity. \\
High-quality example (5): "Clear completion message, smooth transition, consistent visual tone, and a reassuring sense of finish." \\
Low-quality example (1): "Ends abruptly with no confirmation, visual inconsistency, or unresolved states." \\
What to check: \\
- Clear indication the task is complete. \\
- Smoothness of transitions. \\
- Consistency of typography, spacing, and tone. \\
- Absence of unresolved or broken UI states. \\
- Whether the ending reinforces accomplishment. \\
Anchors: 1 = Abrupt, visually inconsistent, or unresolved / 2 = Weak closure or rough visual quality / 3 = Acceptable completion and coherence / 4 = Smooth and cohesive / 5 = Polished, reassuring, and clearly complete.
\end{promptbox}

\paragraph{(6) Specificity}
\begin{promptbox}
Definition: The extent to which the interface visibly incorporates the user's specific situation through wording, examples, or UI choices. This rubric evaluates contextual grounding, not reading level. \\
High-quality example (5): "The prompt and button labels reuse the user's phrase about the late night bug fix and reference the upcoming morning deadline." \\
Low-quality example (1): "Generic prompts that could belong to any user." \\
What to check: \\
- Are the user's phrases reused accurately? \\
- Are contextual constraints reflected? \\
- Is the situation referenced more than once? \\
- Does the interface feel moment-specific? \\
Anchors: 1 = Fully generic / 3 = Topic-level alignment only / 5 = Repeated, concrete contextual grounding.
\end{promptbox}

\paragraph{(7) Understandability}
\begin{promptbox}
Definition: The degree to which instructions use simple, everyday language while reflecting the user's own framing. This rubric evaluates vocabulary level and linguistic accessibility, not layout or personalization depth. \\
High-quality example (5): "Uses the user's own words and explains the step in one short, concrete sentence." \\
Low-quality example (1): "Abstract or technical phrasing disconnected from the user's language." \\
What to check: \\
- Short, concrete sentences. \\
- Everyday vocabulary. \\
- No abstract or technical phrasing. \\
- Clear action in one pass. \\
- Language that mirrors the user's framing. \\
Anchors: 1 = Jargon-heavy and abstract / 3 = Mostly clear but somewhat complex / 5 = Very simple, direct, and grounded in user language.
\end{promptbox}

\end{UIST}


\end{document}